\documentclass[journal]{IEEEtran}

\usepackage{amsmath,amsfonts}
\usepackage{algorithmic}
\usepackage[]{graphicx}
\usepackage{textcomp}
\usepackage{xcolor}
\usepackage{subcaption}
\usepackage{caption}
\usepackage{hyperref}
\usepackage{multirow}

\def\xx{\mathbf{x}}
\def\ww{\mathbf{w}}
\def\hh{\mathbf{h}}
\def\zz{\mathbf{z}}
\def\rr{\mathbf{r}}
\def\bb{\mathbf{b}}
\def\dt{\Delta{t}}
\def\WW{\mathbf{W}}
\def\UU{\mathbf{U}}
\def\RRR{\mathbb{R}}
\def\NNN{\mathbb{N}}

\def\aa{\boldsymbol{\alpha}}
\def\yy{y_\tau}
\begin{document}
% \title[Sensor Network and AI for eSports Analytics]{
% Predicting eSports Athletes' Performance Dynamics Through the Heterogeneous Sensor  Data and Recurrent Neural Network with Input Attention
% % Deep Learning Approach for Modeling Esports Athletes’ Behavior Through Data from Sensors.
% }
% \title[Sensor Network and AI for eSports Analytics]{Big Brother is Sensing You: AI System for Predicting eSports Athletes' Performance by Sensor Data
\title{AI-enabled Prediction of eSports Player Performance Using the Data\\ from Heterogeneous Sensors
%Big Brother is Sensing You: AI System for Predicting eSports Athletes' Performance by Sensor Data
% Other Title Ideas:
% Watch Your Back: Predicting eSports Athletes' Performance Through the Heterogeneous Sensor Data
% Predicting eSports Athletes' Performance Through the Heterogeneous Sensor Data and Recurrent Neural Network
% Deep Learning Approach for Modeling Esports Athletes’ Behavior Through Data from Sensors.
}

\author{
	\IEEEauthorblockN{
		\textsuperscript{}Anton Smerdov,
		\textsuperscript{}Andrey Somov,
		\textsuperscript{}Evgeny Burnaev, Anton Stepanov
		}
	
	\IEEEauthorblockA{
	\textsuperscript{}Skolkovo Institute of Science and Technology, CDISE, Moscow, Russia}
	
	}
% \author{Anton Smerdov, Evgeny Burnaev, Andrey Somov}
% \email{Anton.Smerdov@skoltech.ru, e.burnaev@skoltech.ru, a.somov@skoltech.ru}
% \affiliation{
%   \institution{Skolkovo Institute of Science\\ and Technology}
%   \city{Moscow}
%   \country{Russia}
% }

%\author{Evgeny Burnaev}
%\affiliation{
%  \institution{Skolkovo Institute of Science and Technology}
%  \city{Moscow}
%  \country{Russia}}
%\email{e.burnaev@skoltech.ru}

%\author{Andrey Somov}
%\affiliation{
%  \institution{Skolkovo Institute of Science and Technology}
%  \city{Moscow}
%  \country{Russia}}
%\email{a.somov@skoltech.ru}



\markboth{IEEE Transactions on Instrumentation and Measurement}%
{Smerdov \MakeLowercase{\textit{et al.}}: AI-enabled Prediction of eSports Player Performance Using the Data from Heterogeneous Sensors}

\maketitle

\begin{abstract}
The emerging progress of eSports lacks the tools for ensuring high-quality analytics and training in Pro and amateur eSports teams.
We report on an Artificial Intelligence (AI) enabled solution for predicting the eSports player in-game performance using exclusively the data from sensors. 
For this reason, we collected the physiological, environmental, and the game chair data from Pro and amateur players.
%The recorded data include the Pro and amateur players' physiological signals, the environmental data, and the gamer chair movements. 
The player performance is assessed from the game logs in a multiplayer game for each moment of time using a recurrent neural network. 
%We have trained four predictive models, and selected a recurrent neural network demonstrating the best results. 
We have investigated that attention mechanism improves the generalization of the network
% helps the network perform better
and provides the straightforward feature
importance
% interpretation
as well.
The best model achieves ROC AUC score 0.73. The prediction of the performance of particular player is realized although his data are not utilized in the training set.
%in predicting whether a player not presented in the training set will perform better or worse in the future.
%To the best of our knowledge,
% our work is the first
%this is the first work
%to predict the player in-game performance
%relying exclusively on
% exclusively (?)
% sensors data.
% The proposed solution has a number of promising applications for Pro eSports teams as well as a learning tool for amateur players.
The proposed solution has a number of promising applications for Pro eSports teams and amateur players, such as a learning tool or performance monitoring system.

%as a learning or selection tool.

% to predict player  and found recurrent neural network

% In this work, we propose a sensor network for the data collection in eSports domain. 
% As a target we used the players performance calculated from the in-game replays. Our ultimate goal is to predict the player performance dynamics.

% After the processing data from sensors is provided into the classical machine learning algorithms as well as recurrent neural network. It turned out that the latter performs much better, espesially with the input attention layer added. Attention mechanism helped to improve the generalization of the model and to interpret the feature importance for each moment of time.
% The important findings in our work is that even the raw data from sensors can provide valuable information about the players perfomance dynamics in a first-person shooter game. 
%This knowledge is useful for professional eSports athletes or coaches, as well as eSports enthusiasts considering eSports career in the future.
% We have conducted experiments with 21 participants, including professional eSports athletes

% We also 
\end{abstract}


% \IEEEpeerreviewmaketitle

\IEEEpeerreviewmaketitle

\section{Introduction}

% Одна из фишек рисерча - алгоритмы подходят для всех скиллов, то есть в мотивации можно уделить внимание не только профессиональным геймерам, но и любителям и новичкам, желающим прокачать свои навыки.

% eSports is becoming widely acceptable industry. But still, there is lack of eSports research.
% It's not a surprise that nowadays professional 

eSports is an organized video gaming where the single players or teams compete against each other
% with the aim of achieving
to achieve 
a specific goal by the end of the game. The eSports industry has progressed a lot within the last decade~\cite{esports-2018}: a huge number of professional and amateur teams take part in numerous competitions where the prize pools achieve tens of millions of US dollars. Its global audience has already reached 380 mln. in 2018 and is expected to reach more than 550 mln. in 2021~\cite{newzoo-2018}. eSports industry includes so far a number of promising directions, e.g., streaming, hardware, game development, connectivity, analytics, and training. 

Apart from the growing audience, the number of eSports players and Pro-players (or athletes), the players with a contract, has tremendously grown for the last few years. It made the competition among the  players and teams even harder, attracting extra funding for training process and analytics. The opportunity to win a prize pool playing a favorite game is very tempting for amateur players, and most of them consider a professional eSports career in the future. At the same time, \textit{analytics and training} direction is recognized as the most promising one as it includes the innovative research and business in artificial intelligence, data/video processing, and sensing.

Although eSports is recognised in many countries as sport, it is still in infancy period: there is a lack of training methodologies and widely accepted data analytics tools. It makes it  unclear how to improve the particular game skill except for spending the lion share of time in the game and watching how the popular streamers perform, and participating in trainings. Currently, there is a lack of tools providing feedback about the player performance and advising how to perform better.
It creates a huge potential for the eSports research in order to understanding the factors essential to win in a game. Considering the abundance of data available through the game replays, so-called ’demo’ files, allow for replicating the game and performing fundamental analysis. This kind of analytics is available for both amateur players and professional eSports athletes. 

In terms of prediction and analytics, which is relevant to the research reported in this work, most of the current works in eSports rely exclusively on the in-game data analysis. However,
using only in-game data for estimating the players' performance is a limiting factor for providing helpful feedback to the team and players.
While it can provide the primary information about the gamer's traits and behavior, the huge amount of data from the physical world and captured by sensors~\cite{serious-games} is omitted.
Moreover, sensor data may be more suitable for the eSports domain since models trained on
in-game data only quickly become obsolete when a new patch
is released.
% Sensor data can supplement logs obtained from
% in-game data to provide additional information for predictive
% models and potentially improve their performance.
Information about the player's physiological conditions, e.g. heart rate, muscle activity, and movements can supplement logs obtained from
in-game data to provide additional information for predictive
models and potentially improve their performance.
Multimodal systems utilizing this information have already been
% used with
explored for
audio, photo, and video stimuli \cite{psycho_physical_analysis_multimodal}.
% Similar approaches with multimodal systems has been already explored for audio, photo and video stimuli \cite{psycho_physical_analysis_multimodal}.

In this article, we report on predicting the eSports player performance using the data collected from different sensors and recurrent neural network for data analysis. 
While there is a number of relevant research papers dealing with the prediction of a player skill in general, to the best of our knowledge, there is no research on the estimation of current player performance at a particular moment of time relying using various sensors and the data collected from Pro players. This immediate prediction can provide the instantaneous feedback and can serve as a useful tool for the eSports  team analysts and managers to monitor the current conditions of players. Another practical application is the real-time performance monitoring tool for eSports enthusiasts who want to progress towards the professional level and sign a contract with a professional team.
Since playing a game is a high mental load and stress, we propose to use a multimodal system to record the players' physiological activity (heart rate, muscle activity, eye movement, skin resistance, mouse movement), gaming chair movement, and environmental conditions (temperature, humidity, and $CO_{2}$ level).
This data may help explain variations in gaming performance during the game and to identify which factors affect the performance the most.

Contribution of this work is threefold. (i) Experimental testbed and heterogeneous data collection from various sensors. The dataset is collected in collaboration with a professional eSports team. (ii) Investigation of the optimal neural network architecture for predicting a  player performance and interpreting the obtained results. (iii) In terms of data analysis, we made a special emphasis on the current performance status of the player instead of considering the overall player skill.

This paper is organized as follows: in Section~\ref{related_work} we introduce the reader to the relevant research in the area. Afterwards, we present methods used in this research in Section~\ref{sensor_network}. Experimental results are demonstrated in Section~\ref{results_section}. Finally, we provide concluding remarks in Secion~\ref{conclusions_section}.

%the sensor network architecture with a special emphasis on synchronization. In Section~\ref{data_collection} we focus on the data collection and game scenario. Next, we discuss the data pre-processing and predictive models used in this work in Section~\ref{data_preprocessing} and Section~\ref{alg_section}, respectively. We present the evaluation and experimental results in Section~\ref{evaluation_section} and Section~\ref{results_section}, respectively. We discuss the feature importance in Section~\ref{feature_importance_section} and provide concluding remarks in Sections~\ref{discussion_section}~and~\ref{conclusions_section}.
 
\section{Related Work} \label{related_work}
Wearable sensors and body sensor networks have been widely applied for assessing a human behaviour and activity recognition in many areas~\cite{gender-recognition}. However, this approach has not been extensively used for assessing eSports players: typically in-game data or data collected from computer keyboard and mouse have been analyzed so far. Due to this limitation the performance evaluation methods are limited as well.

This section is therefore divided into two parts: first, we overview relevant research in terms of data collection and activity recognition and, second, we discuss recent research on performance evaluation methods.

%There is not much prior research regarding utilizing sensors data to predict eSports player behavior.
%However, there is much work
%either
%using sensors data to analyze people's behavior in other domains,
% as well as research
%either
%analyzing gamer behavior by in-game data instead of sensors data.
%Thus, prior research is naturally categorized by the type of data used
% to analyze the behavior of a player or other person.
%for analysis:

%\begin{enumerate}
%    \item Sensor data
    % for skill assessment
%    in eSports \label{sensor_esports}
%    \item In-game data
    % for skill assessment
%    in eSports
%    \item Sensor data
    % for skill assessment
%    in other domains
%\end{enumerate}

%We are mostly interested in prior work in
% \ref{sensor_esports}, 
%the first category; however, research in the other two categories is relevant as well, because it differs only by the type of data used or by the domain.

\subsection{Activity Recognition Using Sensors}
Indeed, there is a lack of prior research  utilizing sensors data to predict eSports player behavior. It happens due to young research domain in eSports. Recent research on using sensor data in eSports is limited to predicting the overall player skill or finding simple dependencies in the data.

% There is some prior work on understanding physiological behavior in eSports.
In \cite{heart_rate_correlation_esports} authors investigate the correlations between psychophysiological arousal (heart rate, electrodermal activity) and self-reported player experience in a first-person shooter game.
% Similar research \cite{lol_stress} is also done MOBA genre.
Similar research about the relation between player stress and game experience has been investigated in the MOBA genre \cite{lol_stress}.
% in  has also been investigated  is also done .
The connection between the gaze and player skill is investigated in \cite{visual_fixations}. Mouse and keyboard data is a natural source of information about the players. Its relation with player performance in first-person shooters is covered in \cite{input_skill_prediction} for Red Eclipse and \cite{esports_gaze_mouse_1} for Counter-Strike: Global Offensive (CS:GO).
Player performance can also be predicted by activity on a chair during the game \cite{smart_chair_wf_iot} or in reaction to key game events \cite{smart_chair_iop}.

However, there is extensive research work carried out on data collection and activity recognition in other applications including sports, medicine, and daily activity monitoring. In sports, wearable sensing systems are used for detection and classification of training exercises for goalkeepers~\cite{soccer_detection_classification}, assessing header skills in soccer~\cite{soccer-2020}.
%smart shoes are designed for visualizing the force and centre of pressure of player kicks~\cite{WEIZMAN2015}
Also, wearable systems were designed to classify tricks in skateboarding~\cite{GROH2017}, classify popular swimming styles using  sensors~\cite{swimming}, and other activities in sports~\cite{horse-sport}. In terms of daily activity monitoring and medical applications, they have been studied for nearly three decades with the use of wearable sensors. Many medical studies deal with the investigation of human gait, for example, for patients with the Parkinson's disease~\cite{parkinson}.
%and for patients during the rehabilitation after surgery~\cite{haladjian-2018-knee}. 
%Significant progress in mobile sensing and AI~\cite{sensys-2019-heart}, resulted in appearance of generic mobile sensing systems for daily activity inference~\cite{sensys-2019-activity}. Another promising approach for nonintrusive human behaviour montoring is application of earables~\cite{Earables}, an in-ear multisensory stereo device. We note here, that deploying deep learning into mobile and embedded devices~\cite{Squeezing} opens up wide vista for numerous activity recognition tasks based on data collected from wearable sensors, video cameras, and earables.   

\subsection{Performance Evaluation Methods}

Most of the current research in eSports analytics  relies exclusively on the in-game data collection and further analysis. It has been shown that information about kills, deaths, and other game events can help predict a game outcome in Multiplayer Online Battle Arena (MOBA) discipline for Dota 2~ \cite{dota_match_prediction_1}, League of Legends \cite{lol_match_prediction}, and Rocket League \cite{rocket_league_match_outcome}.
% It's shown that in  genre match outcome can be predicted by information about kills, deaths and other in-game statistics for Dota 2 \cite{dota_match_prediction_1} and League of Legends  \cite{lol_match_prediction}.
% % That approach is for \cite{dota_match_prediction_1}.
% % Similar techniques work for another MOBA game League of Legends \cite{lol_match_prediction}.
Another opportunity to predict the game outcome in the MOBA related disciplines is based on the  features extracted from players' match history, as well as in-game statistics \cite{dota_match_prediction_0}.
Players match history can also be used to create a rating system for predicting the matches outcome in the First-Person Shooter (FPS) genre  \cite{csgo_dota_match_prediction}. 

% Other Relevant research activities in eSports include the predictions based on drafting and matchmaking.
As noticed earlier, the in-game data in eSports is widely used for analytic studies in the area.
Drachen et al. consider clustering a player behavior to learn the optimal team compositions for League of Legends discipline to develop a set
of descriptive play style groupings~ \cite{Drachen2012Guns}.
% feeding these into a prediction model for win/loss outcomes.
% The authors have demonstrated that match win/loss conditions can be predicted from team composition-based features with accuracies around 70\% depending on the classification method used.
Research by Gao et al.~\cite{gao-2013} targets the identification of the heroes that players are controlling and the role they take.
The authors have used classical machine learning algorithms trained on game data to predict a hero ID and one of three roles
and achieve the accuracy ranging from 73\% to 89\% which depends on features and targets used.
Eggert et al.~\cite{eggert2015classification} has continued
the work by Gao et al.~\cite{gao-2013} and applied the supervised machine learning to classify the behavior of DOTA players in terms of hero roles and playstyles. 
Martens et al. \cite{martens2015toxicity} have proposed to predict a winning team analyzing the toxicity of in-game chat.  In \cite{blowout_matches} authors used pre-match features to predict the outcome and analyze blowout matches (when one team outscores another by a large margin).
% Toxic behavior in eSports is another popular research topic.
% For addressing this challenge, the authors in \cite{martens2015toxicity} proposed
% to detect toxic players with
% linear SVM trained on features extracted from
% a team chat
% in Dota 2.
% Authors developed a methodology to annotate frequently used
% expressions in written chat communication of Multiplayer
% Online Games to detect toxicity. The approach was based on contextual information to distinguish simple swearing from deliberate insults.
% % Player and team profiling
% % Выведение закономерностей в геймплее и предсказание реакций на те или иные события.
% These models are usually based on in-game and biometric players data that is built with the following mathematical machinery: regression models and least-square methods (linear, non-linear, and logistic) focused on establishing a linear relationship between considered variables minimizing the sum of squares; Decision trees and gradient boosting. Search the space of future actions and build trees of possible action sequences, often in an adversarial setting.
The research reported in 
\cite{cornforth2015cluster} describes the cluster evaluation, description, and interpretation for player profiles in Minecraft.
% based on log files on a Minecraft game server.
%% Calculated variables were extracted from these logs in order to characterize players. % WHAT (1)
%% Using circular statistics, the authors show how measures can be extracted that enable players to be described by the mean and standard deviation of the time that they interacted with the server. % WHAT (2)
%% Feature selection is accomplished using a correlation study of variables extracted from the log data. %
% The authors extract features about the players' interaction with  circular statistics and then select features 
%% This process favored a small number of the features, as judged by the results of clustering. % WHAT (4)
The authors state that automated clustering methods based on game interaction features help identify the real player communities in Minecraft.
%% Cluster evaluation, description, and interpretation techniques are applied to provide further insight into distinct behavioral characteristics, leading to a determination of the quality of clusters, using the Silhouette Width measure. % Not needed IMO
% Together with the progress in eSports, there are still so-called 'toxic' players aimed at destructive actions in the game and game community.

% In \cite{watanabe2017toward} decision three, RandomForest, Bayesian Network, SVM were applied to data acquired from match replays of League of Legends which included pings in the actions. The  weighted average scores of Precision, Recall, and F-Measure in 10-fold cross-validation for classifiers is 90 and above.
% In \cite{kwak2015exploring} the authors  explored cyberbullying and other toxic behavior in team competition online games. They used a dataset of over
% 10 million player reports on 1.46 million toxic players along with the corresponding crowdsourced decisions. 

%\subsection{Sensor Data
% for Skill Assessment
%in Other Domains}

In many domains the skills and performance can be assessed and/or predicted based on sensors data~\cite{combat-sport}. In sport, the data from the Inertial Measurement Unit (IMU) can be helpful for estimating volleyball athlete skill \cite{volleyvall_skill_sensors}. Tennis player performance can be assessed from the IMU data on the hand and chest \cite{tennis_skill_0}, or on the waist, leg, and hand \cite{tennis_skill_1}.
Similar techniques have been investigated for skill estimation in 
%Australian football \cite{sport_activities_skill_sensors_0},
soccer \cite{soccer_detection_classification}, climbing  \cite{climb_skill_sensors}, golf \cite{sport_activities_skill_sensors_1}, gym exercising \cite{ mobile_exercise_assessment_1}, and alpine skiing \cite{alpine_skiing_skill}.

Another popular domain for skill assessment based on the sensor data is surgery. In~\cite{surgery_skill_sensors_0} authors use IMU data to create a skill quantification framework for surgeons.
% report on a data-driven  based on.
% In addition to motion data
% The research work by
Ershad et al. \cite{surgery_skill_sensors_1} have shown the connection between the surgeon skill and behavior information collected from IMU.
Ahmidi et al. \cite{surgery_skill_sensors_2_eye_tracking} developed a system using motion and eye-tracking data for surgical task and skill prediction. The authors have used hidden Markov models \cite{hidden_markov_models} to represent surgeon state which is similar to the method proposed in this work in Section~III.
% condition can be represented by some hidden state.
The connection between the surgeon actions and pressure sensors data has been investigated in \cite{surgery_skill_sensors_3_pressure_sensors}.

Physiological data have also been used for predicting skill level and skill acquisition
% by data collected during
in
working activities, such as mold polishing \cite{worker_skills_mold_polishing} and clay kneading \cite{worker_skills_clay_kneading}, as well as dancing~\cite{dance_skill}.

In this research, we use a number of wearable and local unobtrusive sensors used for data collection during the game session and further data analysis. It was carried out with respect to the players needs. 

\section{Methods} \label{sensor_network}
\begin{figure}[b]
    \includegraphics[width=1.0\linewidth]{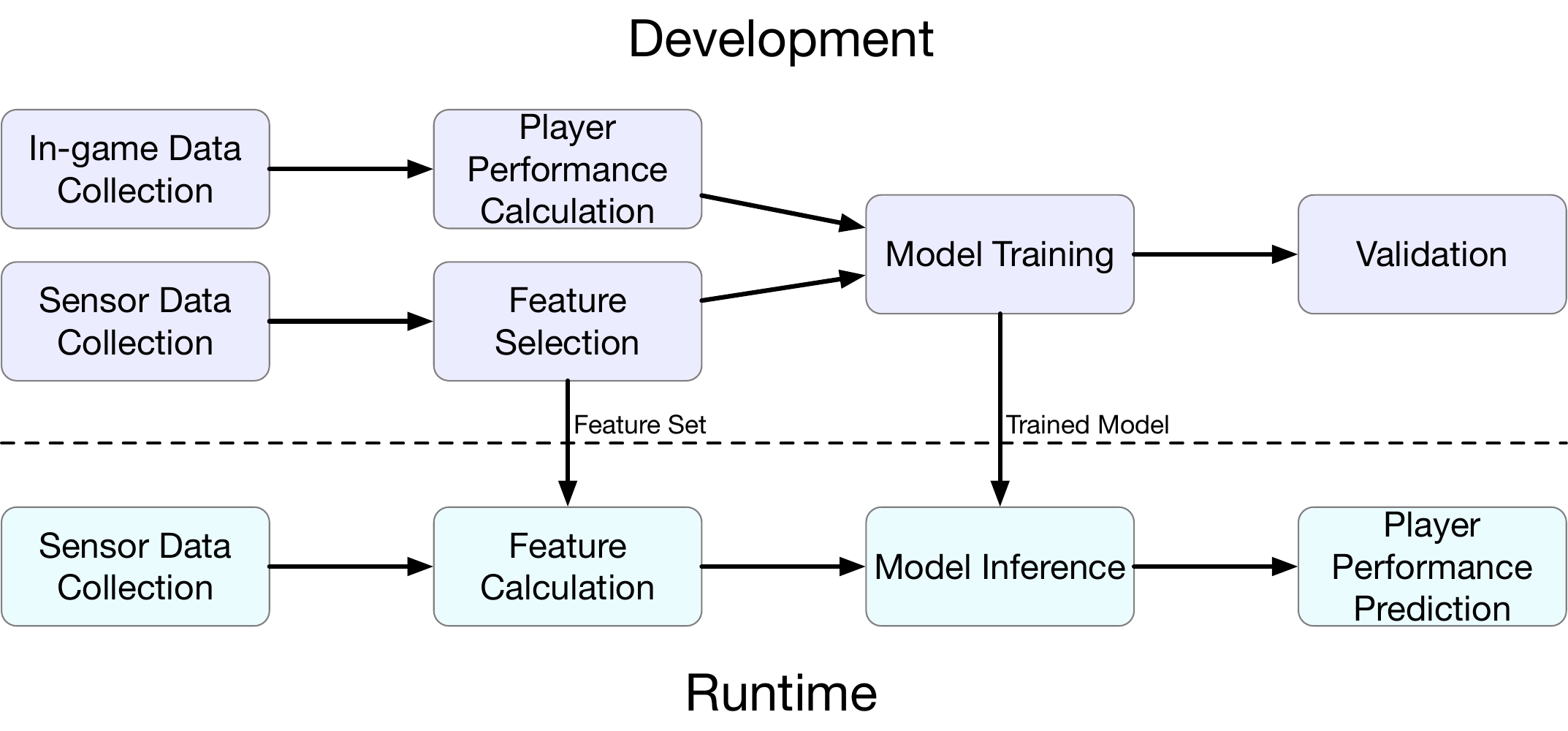}
    \caption{An overview of prediction system during the development and runtime phases.}
    \label{architecture-1}
\end{figure}

In this section, we describe the sensors used in this research, data collection procedure, data pre-processing, and data analysis helping predict the players performance. In Figure~\ref{architecture-1} we present an overview of prediction system.

\subsection{Sensors}

\begin{figure}[!]
    \includegraphics[width=1.0\linewidth]{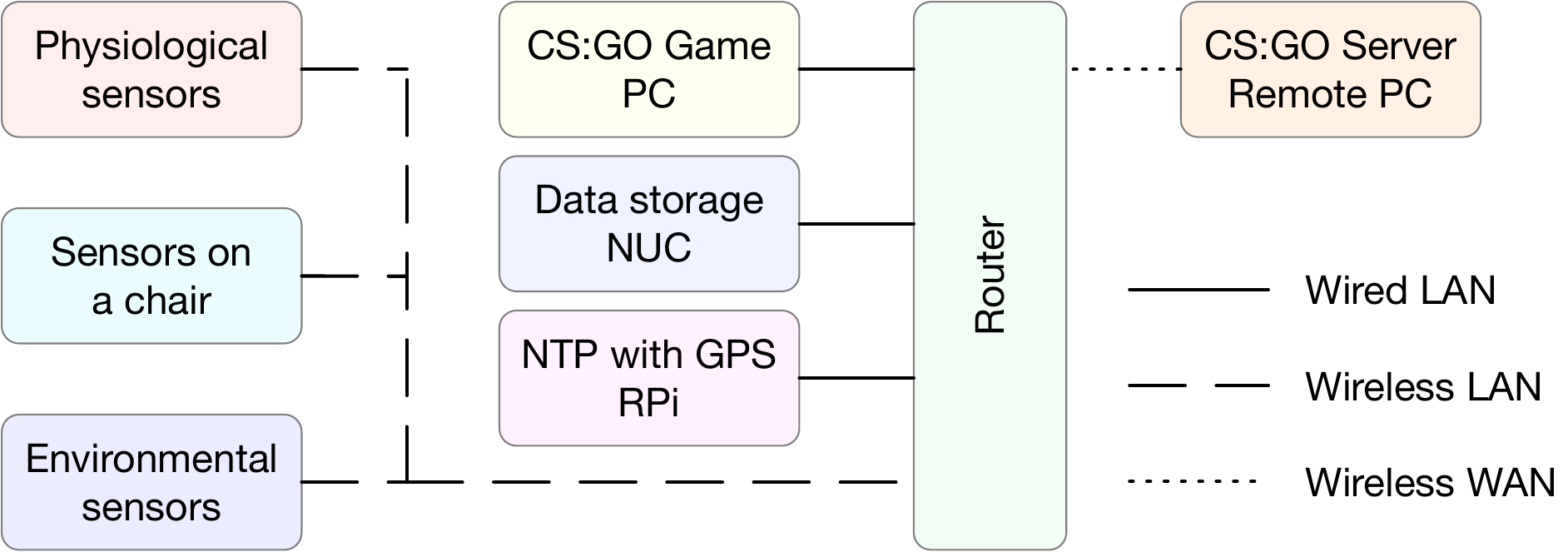}
    \caption{Sensor network architecture.}
    \label{architecture}
\end{figure}

In our work, we use three groups of sensors: \textit{physiological} sensors, sensors integrated into a \textit{game chair}, and \textit{environmental} sensors. The sensor network architecture is shown in Figure~\ref{architecture}. The list of sensors used, their locations, and sampling rates are presented in Table~\ref{sampling_rates}. Further issues associated with sampling rates are discussed in Section \ref{time_series_resampling}.

\begin{table*}[h]
\centering
\caption{Sensors location and sampling rates.}
\label{sampling_rates}
% \begin{tabular}{|p{3cm}|p{5cm}|p{3.2cm}|p{2.4cm}|}
\begin{tabular}{|p{3cm}|p{5cm}|p{3.2cm}|l|}
\hline
\textbf{Sensor Group} & \textbf{Sensor} &  \textbf{Location} & \textbf{Sampling Rate} \\ \hline
\multirow{5}{\linewidth}[-0.0em]{Physiological Sensors} & Electromyography sensor \cite{emg_sensor} & Elbow & 70 Hz\\ \cline{2-4}
& GSR sensor \cite{gsr_sensor} & Fingers & 70 Hz \\ \cline{2-4} 
& Heart rate monitor \cite{heart_rate_monitor} & Chest & 3 Hz\\ \cline{2-4} 
& Eye tracker & Under the PC monitor & 90 Hz \\ \cline{2-4} 
& Mouse & Hand & 250 Hz \\ \hline
Sensors on a Chair  & 3-axis accelerometer and 3-axis gyroscope \cite{mpu9250} & The bottom of the chair & 100 Hz\\ \hline
\multirow{3}{\linewidth}{Environmental Sensors} & CO$_2$ sensor \cite{co2_sensor} & \multirow{3}{\linewidth}{Environment, $\approx$1m from the player} & 0.2 Hz \\ \cline{2-2} \cline{4-4}
& Temperature sensor \cite{temperature_sensor} & & 5 Hz \\ \cline{2-2} \cline{4-4}
& Humidity sensor \cite{humidity_sensor} & & 5 Hz \\ \hline
\end{tabular}
\end{table*}

\begin{samepage}
Physiological data recorded: % rewrite
\begin{itemize}
    \item \textbf{Electromyography} (EMG) data as an indicator of muscle activity. EMG data are related to physiological tension affecting player's current state \cite{emg_tension}.
    \item \textbf{Heart rate} data received by a heart rate monitor on a chest. High values correspond to mental stress and arousal \cite{heart_rate_stress}, which might affect the rationality of player decisions.
    \item \textbf{Electrodermal activity} (GSR) or skin resistance data as a measure of person arousal \cite{gsr_arousal}. This value is also connected with the stress level.
    \item \textbf{Eye tracker} data of player gaze position on a monitor in pixel values. The player must check the minimap and other indicators on the screen to have relevant information about the game and, thus, make effective decisions \cite{visual_fixations}. 
    \item \textbf{Mouse movements} captured by a custom python script as a measure of the intensity of a player input. This data is an  indirect indicator of the hand movement activity as well as the player skill \cite{input_skill_prediction}.
\end{itemize}

\end{samepage}

% 3-axial accelerometer for capturing 
% 3-axial gyroscope for recording rotational movement data.
The sensors integrated into a game chair are presented by a 3-axial accelerometer and a 3-axial gyroscope. We illustrate axes orientation for the chair in Figure~\ref{chair_axes}. Recorded data includes:

\begin{figure}[b]
    \centering
    \includegraphics[width=0.45\linewidth]{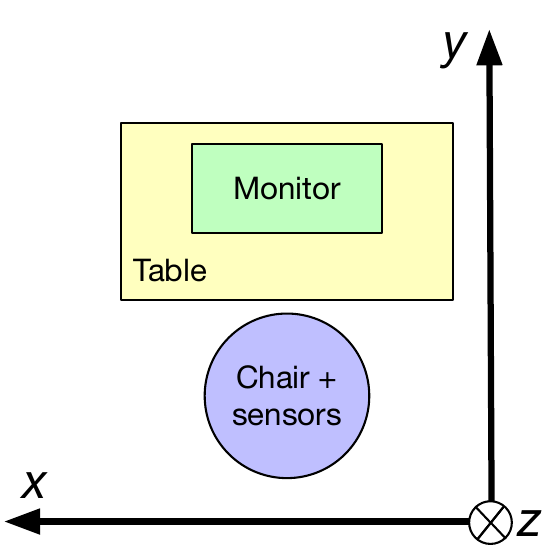}
    \caption{Axes orientation for the chair.}
    \label{chair_axes}
\end{figure}

\begin{itemize}
    \item \textbf{Linear acceleration} of a chair. It captures the player movements to the game table, parallel to it, chair height changes, and  small oscillations possible in stress conditions. Behavior on a chair is connected with the player skill \cite{smart_chair_iop, smart_chair_wf_iot}.
    \item \textbf{Angular velocity} of a chair. This data provides the information about the person's wiggling and spinnings on a chair.
\end{itemize}

Environmental data recorded:
\begin{itemize}
    \item \textbf{$CO_2$ level}. High $CO_2$ level results in the reduction of cognitive abilities \cite{co2_performance}, thus directly affect the gaming process.
    \item \textbf{Relative humidity}. High level of relative humidity results in the reduction of  neurobehavioral performance \cite{humidity_performance}
    %  performance of neurobehavioral tests was lower at higher relative humidity 
    \item \textbf{Environmental temperature}. Too warm conditions may affect the human performance \cite{temperature_performance}.
\end{itemize}

\subsection{Sensor Network and Synchronization}

Apart from a number of heterogeneous sensors, the sensor network has a dedicated storage server (based on Intel NUC PC), gamer PC (a high-speed Intel I7 PC with the DDR4 memory,
and an advanced GPU card). For synchronization reasons, we have an NTP server
with GPS/PPS support. A high-speed wireless router connects all the devices in the network. PC with strict requirements to ping value (gamer PC, NTP server) has a wired connection to the router (LAN). The sensors have a wireless connection to the network as they are placed near the eSports athletes (WLAN). The router has a low latency connection to the Internet (WAN).
%\subsection{Synchronization}
Proper synchronization of the sensors and gamer PC is essential for further data collection and analysis.

\subsubsection{NTP Server}
At present, there are many options for building time-synchronous systems for industrial applications, e.g., TSN by NI\footnote{\url{http://www.ni.com/white-paper/54730/en/}}.
At the same time, there must be a reasonable balance between the cost of synchronization solution and its accuracy. The cost of most industrial solutions is high, preventing them from integrating into the player's PC. It happens since the desktop computers are primarily selected according to the 3D games performance criteria and do not have specific hardware devices on board. That is why we decided to realize the  synchronization on a single NTP server.
%\footnote{\url{https://tools.ietf.org/html/rfc5905}}. 
A reliable and always available server which could be located close enough and characterized by the minimum delay in transmitting the packets over the network is vital for our sensor network. 

A single-board computer Raspberry PI 3B
%\footnote{\url{https://www.raspberrypi.org/products/raspberry-pi-3-model-b/}} 
was selected as a server, and a GPS signal was used as a source of reasonably accurate time. The signal from the satellite was received by a separate module based on the MTK MT3333 chipset and having the UART interface as well as supporting the PPS signal
%\footnote{\url{http://pos.mgb-tech.com/insightpps/}}. 
on Raspbian Stretch OS,
%\footnote{\url{https://downloads.raspberrypi.org/raspbian_lite/images/raspbian_lite-2018-03-14/2018-03-13-raspbian-stretch-lite.zip}}
GPS support packages (gpsd, gpsd -clients, pps-tools) and Chrony time server
%\footnote{\url{https://chrony.tuxfamily.org/comparison.html}}
was installed. Raspberry PI was located near the window for better satellite signal reception and connected to the local area network via a wired interface. The presence of a dedicated PPS signal acquired by a separate IO pin (GPIO) Raspberry PI made it possible to ensure time accuracy in the range of $10^{-5}-10^{-6}$ s (time accuracy of $1 - 10$ us). 

\subsubsection{Sensors}
The sensors in our network are deployed on Raspberry PI (RPi). The broadcast network "sync" command was sent to the sensors prior to measurements. After the  command reception, a custom made script synchronized the local time to the local NTP server (Stratum 1) time on each RPI. Feedback status with the current time difference was also reported from every RPI to the local data storage PC. In this case, all RPI were synchronized before the  measurement procedure starts. Time drift of local RPI time was measured: it is in the range $10-20$ $ms$ per hour. In this case, the sync command was repeated every $10$ minutes. This allows us to have synchronized sensors all the time. 

\subsubsection{Gamer PC Synchronization}
%Gamer PC also has several local sensors (mouse and keyboard loggers, eye-tracker and etc.). 
Performing the time synchronization on a gamer PC was another significant issue. The players used MS Windows OS on their PCs which do not provide the  accurate time to the user by default (you can check how accurate the clock on your PC is at \url{www.time.is}). The default settings in Windows 7/8/10 allow the users to synchronize time with the NTP server only once a week. At the same time, the average time drift of the clock is 50 $ms$ per hour and even more for an ordinary PC according to our measurements.
In MS Windows 10 OS build 1607 and newer, there is a way to reduce the synchronization period and get significantly higher time accuracy by setting the registry.

Then Windows Time Service should be switched to Auto (always loaded after the PC starts) start mode.

The accuracy of the clock within 1 $ms$ requires to meet a number of conditions\footnote{\url{https://docs.microsoft.com/en-us/windows-server/networking/windows-time-service/support-boundary}}. In our experiment (taking into account the local time server Stratum 1 based on RPI), all the  requirements were met with the exception of the ping value (it was $<1$ ms, instead of the required value $<0.1$ ms). However, it allows  us to achieve the necessary synchronization accuracy.

In the case of proper registry settings after some time, the drift is compensated by the internal Windows algorithms, and the clocks become synchronous with the time server (within $2-3$ $ms$ accuracy).

Upon synchronizing the hardware in the network, we start the data collection procedure.

\subsection{Data Collection} 
\label{data_collection}

\begin{figure}[t]
\center{
    \includegraphics[width=0.8\linewidth]{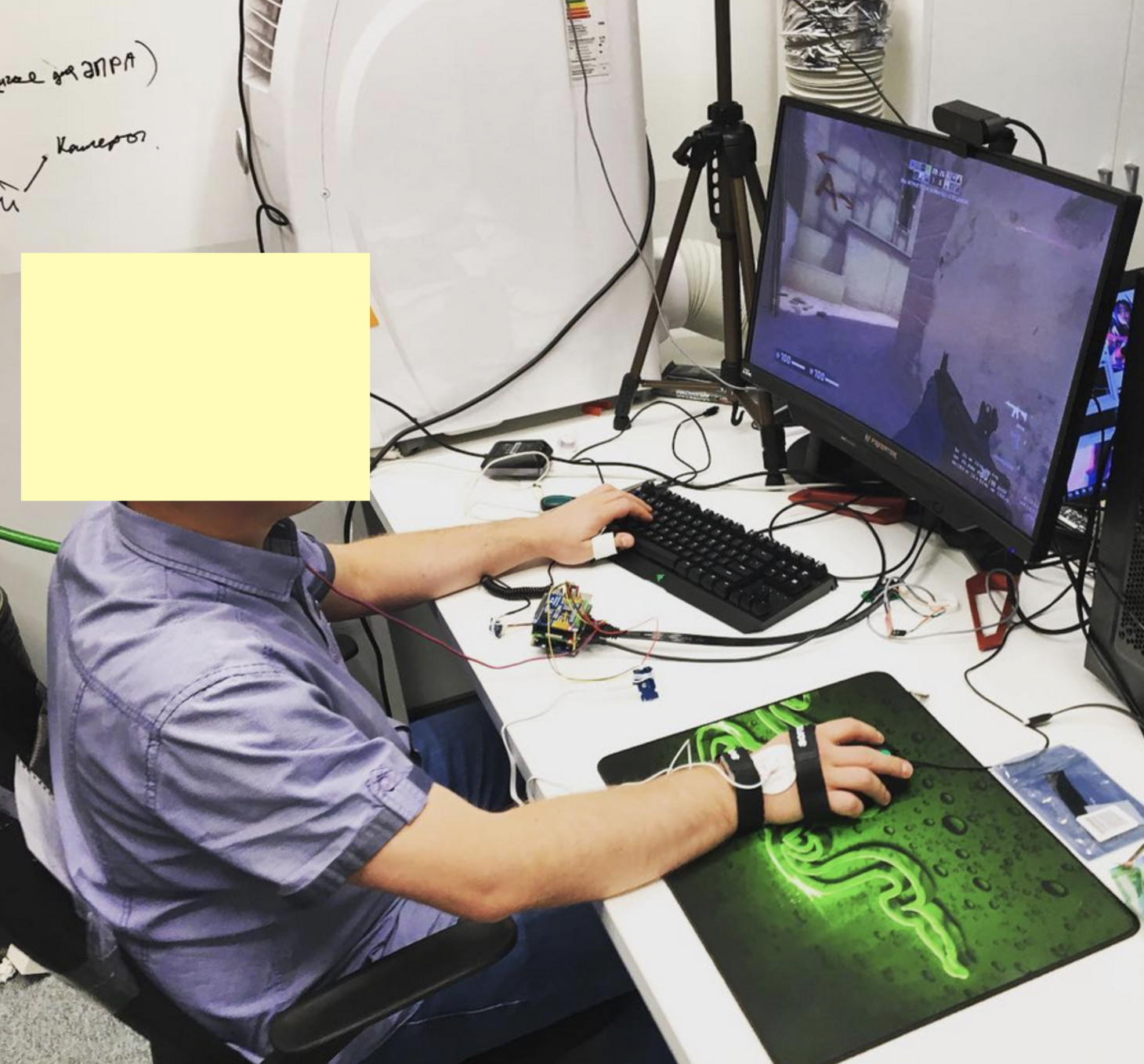}}
    \caption{Experimental testbed.}
    \label{testbed}
\end{figure}

We invited 21 participants to play FPS  Counter-Strike: Global Offensive (CS:GO) for 30-35 minutes. We note here that six pro-players took part in this experiment. All the participants were informed about the project and the experimental details. Every participant signed a written consent form which allowed recording physiological and in-game data. Then players were equipped with the sensors for data collection. We did not receive any complaints about the uncomfortable gaming experience because of the sensors. The experimental  testbed is snown in Figure~\ref{testbed}.

% \begin{figure}[!ht]
% \centering
% \begin{minipage}{.4\textwidth}
%   \centering
%   \includegraphics[height=4.5cm]{pic/testbed_anon.png}
%     \caption{Experimental testbed.}
%     \label{testbed}
% \end{minipage}%
% \begin{minipage}{.6\textwidth}
%   \centering
%     \includegraphics[height=5cm]{pic/screen-interface.png}
%     \caption{CS:GO screen interface. Most of the time the pro-players watch at the aiming crosshair.}
%     \label{screen}
% \end{minipage}
% \end{figure}

\begin{figure}[t]
\center{
    \includegraphics[width=0.8\linewidth]{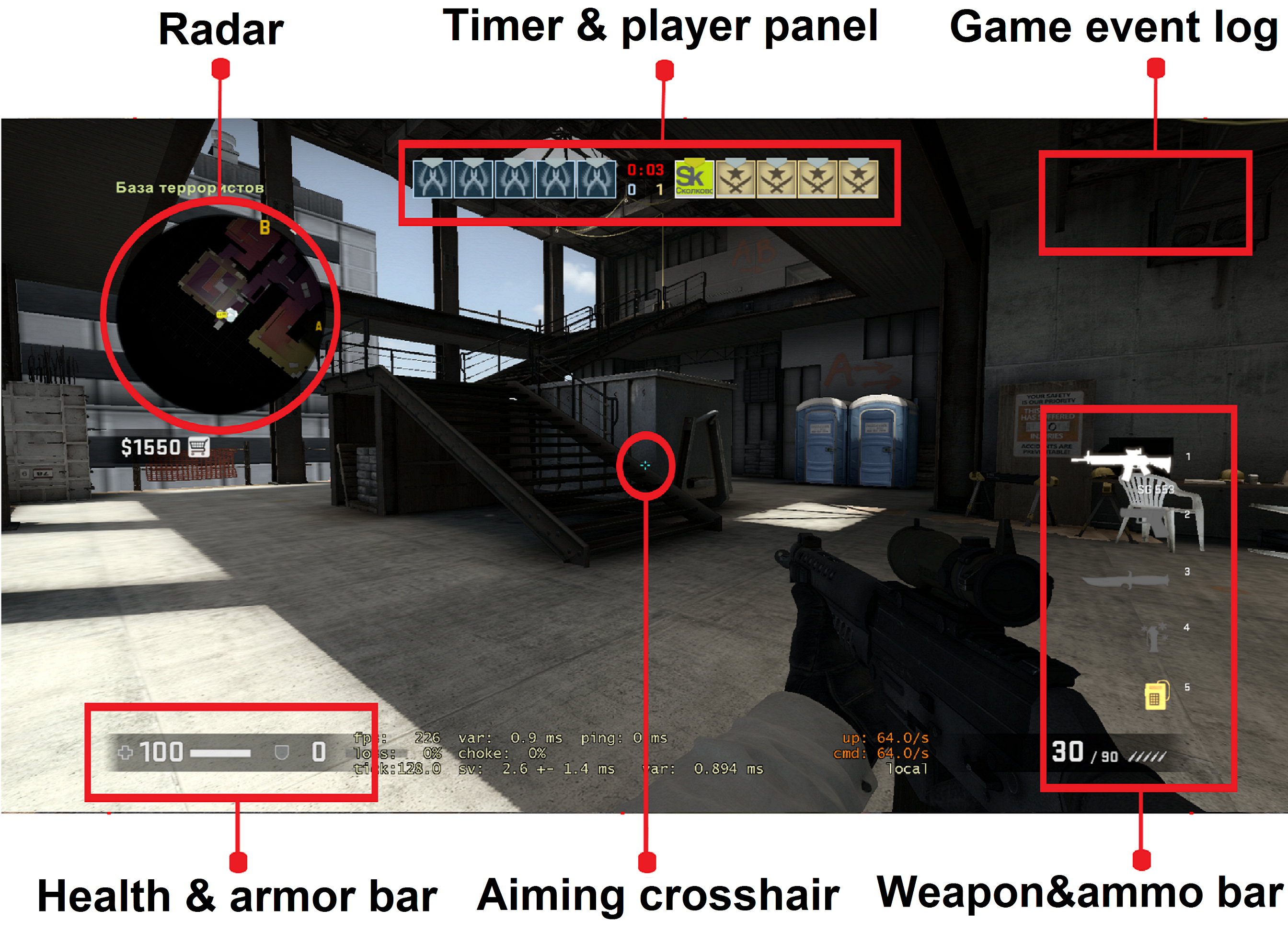}}
    \caption{CS:GO screen interface. Most of the time the players watch at the aiming crosshair.}
    \label{screen}
\end{figure}

Players needed to play \textit{Deathmatch} mode of CS:GO. The example of the game screen interface is shown in Figure~\ref{screen}.
In this mode the goal of each player is to achieve as many kills of other players as possible and to minimize the number of their own deaths. When the player is killed it immediately respawns in a random  location in the game.
This mode is often used by eSports players in their training routine. 
% As usual for Counter-Strike, the counter-terrorist team is opposed to the terrorist team. The 
% % The key feature of the mode is that the terrorist team consists of two players who typically play in a defensive manner and should defend a bomb from the opposite team. The counter-terrorist team is made up of 3 players who know the location of the bomb and have a goal to defuse it or to kill all terrorist players. One match consists of 12 rounds with approximately 40 seconds duration each. The match is played without any breaks between the rounds.
% The participants did not have any breaks between consecutive rounds of the game.
After the game had been
% ended,
finished,
we saved the replays for the future game events extraction.

\begin{figure*}[t]
    % \begin{subfigure}[b]{\textwidth}
    \includegraphics[width=\linewidth]{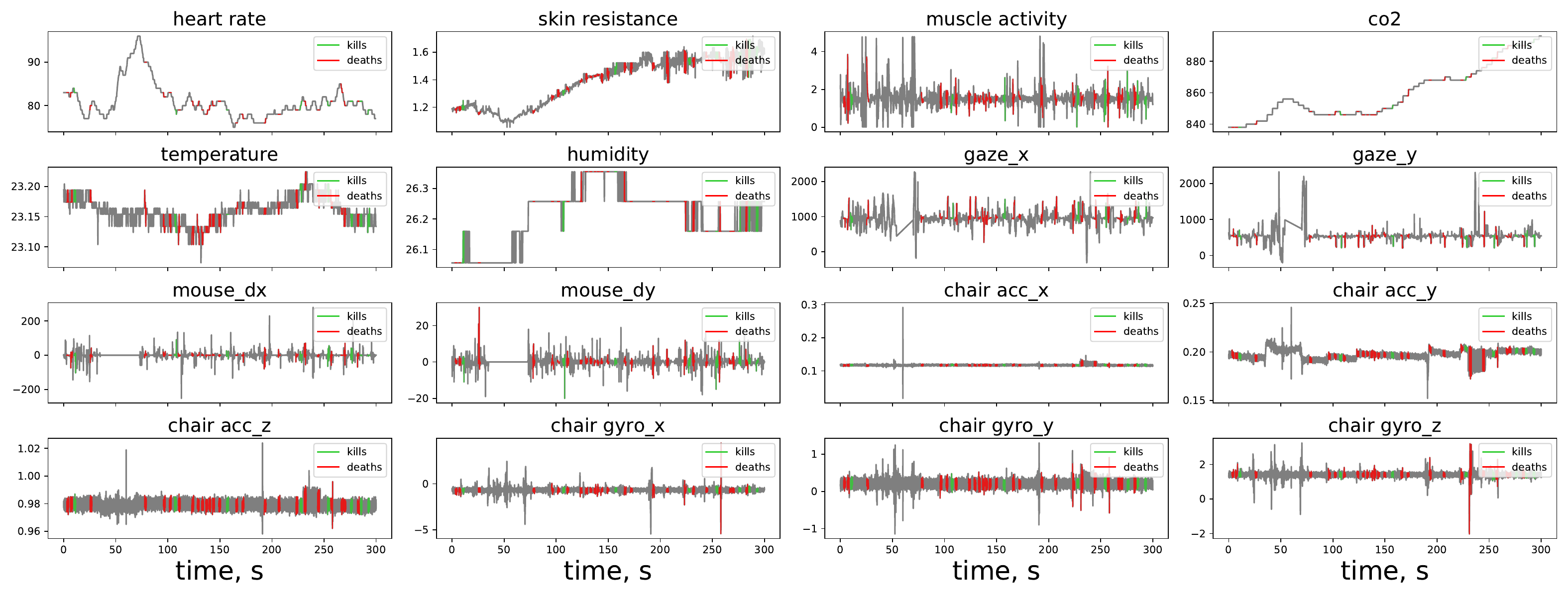}
    \caption{5-minutes data samples for one player. Green and red coloring correspond to \textit{kill} and \textit{death} events, respectively.}
    \label{data_samples}
\end{figure*}

Collected data samples for the 5-minutes intervals for two players are shown in Figure~ \ref{data_samples}. For the reader convenience, we color the intervals in the 1-second vicinity of kill and death events.

It is clear that for the data from some physiological indicators, e.g. skin resistance or heart rate, there are global and local trends, which might align with changes in the player's efficiency in the game.
Another point is that the players usually do not move a lot on a gaming chair. However, they can change their posture from time to time, and this event is captured by the IMU on the chair and also might be connected with the game events and player performance.

\subsection{Data Pre-processing} \label{data_preprocessing}

To get rid of the noise and occasional outliers, we have clipped all the data by 0.5 and 99.5 percentiles and smoothed the data by 100 $ms$ moving window.
% For additional smoothing we 
We have also reparametrized gaze, mouse, and muscle activity signals. Mouse signal has been converted from $x$ and $y$ increments to Euclidian distance passed to reflect the mouse speed; gaze data has been transformed from $x$ and $y$ coordinates to Euclidian distance passed; muscle activity signal has been changed to L1-distance to the reference level for the player in order to represent the intensity of muscle tension.
For 3.7\% data missed we have used linear interpolation to fill in the unknown values
% median values for the player
since it provides a stable and accurate
approximation.
% % estimation.

% \subsection{Time series resampling} \label{time_series_resampling}
\subsubsection{Sensor Data Resampling} \label{time_series_resampling}

In order to predict the player performance at each moment of time, it is convenient to resample the data from all the sensors to the common sampling rate. This helps apply the proposed data analysis  for discrete time series predictions, such as hidden Markov models~\cite{hidden_markov_models} or recurrent neural networks~\cite{rnn_review}.

% Since we measured movements along $x$ and $y$ axes for mouse and gaze independently, we transformed this data to Euclidian distance for each measurement before the resampling to extract more meaningful signal.

However, data from different sensors has different underlying nature and should be resampled accordingly. While it is reasonable to average the data within a time step interval for heart rate, skin resistance, muscle activity, environmental data, chair acceleration and rotation, averaging is not applicable for the gaze movement and mouse movement data. The reason behind it is that we are interested in the total distance passed within the time step instead of the average distance passed per measurement. The total distance does not depend on a number of samples, but only on their sum.

% As an example, if a person moves the mouse for 100 pixels for two consecutive time steps with 2 and 4 samples in each, that doesn't mean that mouse speed for the first case is two times higher than in the second case. The mouse and gaze speed doesn't depend on number of samples in the time step, so we must apply summation instead of averaging for them.

Resampling introduces an important hyperparameter \textit{time step}. Throughout the manuscript we  refer to it is as $\dt \in \RRR$. Big time step values, e.g. 5 minutes, are not meaningful for our problem since we need to extract the relevant information about the player. On the other hand, too small timestamp, e.g 0.1 $s$, may lead to an excessive number of observations and noisier data. Indeed, the resampling time step should not be smaller than the time between the measurements for the majority of sensors.

% After the resampling,
After converting the sampling rates to the common value we obtained a 15-dimensional feature vector for each moment of time. Further in the paper we will refer to this feature vector as $\xx(t)\in\RRR^{n}$. Its components are described in Table~\ref{features}. 

\begin{table}[t]
\caption{Feature vector components.}
\label{features}
\begin{tabular}{|l|l|}
\hline
\textbf{Feature Group}                      & \textbf{Feature Name}                  \\ \hline
\multirow{6}{*}{Physiological Activity} & Heart Rate                    \\ \cline{2-2} 
                                   & Muscle Activity               \\ \cline{2-2} 
                                   & Skin Resistance               \\ \cline{2-2} 
                                   & Gaze Movement                 \\ \cline{2-2} 
                                   & Mouse Movement                \\ \cline{2-2} 
                                   & Mouse Scroll                  \\ \hline
\multirow{2}{*}{Chair Movement}    & Linear Accelerations along $x,y,z$ axes  \\ \cline{2-2} 
                                   & Angular velocity along $x,y,z$ axes \\ \hline
\multirow{3}{*}{Environment}       & CO$_2$ level                  \\ \cline{2-2} 
                                   & Temperature                   \\ \cline{2-2} 
                                   & Humidity                      \\ \hline
\end{tabular}
\end{table}

\subsection{Player Performance Evaluation} \label{player_performance_evaluation}

There is no generic player effectiveness metric for the majority of eSports disciplines. The most popular evaluation metric for FPS and MOBA games is Kill Death Ratio (KDR)~\cite{kdr_performance}. It equals the number of kills divided by the number of deaths for the time interval. If KDR > 1, it means that the player performs well, or, at least, better than some players on a game server. Otherwise, the player most likely performs bad compared to other players.

KDR takes values from $0$ to $+\infty$ which is not a clear range for prediction. When the player performs very well and has many kills and few deaths, KDR is fluctuating drastically because of division by a small number. In opposite, if there are many deaths and few kills, KDR is around 0 and changes slowly. This drastic inconsistency in the  target creates difficulties for training machine learning algorithms.
One possible solution could be to apply logarithm to KDR, but this does not solve the issue with the  scale, because logarithm takes the values from $-\infty$ to $+\infty$.

We propose a more numerically stable target value which equals the proportion of kills for the player. More precisely,
\begin{gather}
p_{\tau}(t) = \frac{k_{\tau}(t)}{k_{\tau}(t) + d_{\tau}(t)}, \label{proportion}\\
k_{\tau}(t) = K(t + \tau) - K(t), \label{kills}\\
d_{\tau}(t) = D(t + \tau) - D(t), \label{deaths}
\end{gather}
where $p_{\tau}(t)\in\RRR$ is the proportion or performance and equals the proportion of kills for a player at the moment $t\in\RRR$ considering the  kills and death 
in the next $\tau\in\RRR$ seconds.
In other words, it is the ratio of kills in the next  $\tau$ seconds.
$K(t)\in\RRR$ and $D(t)\in\RRR$ are the total number of kills and deaths at the moment $t$; therefore $k_{\tau}(t)\in\RRR$ and $d_{\tau}(t)\in\RRR$ equals the number of kills and deaths within the interval $[t, t + \tau]$.
$p_{\tau}(t)$ varies from 0 to 1 and has higher values for well-performing players. Bounding by 0 and 1 helps efficiently train the machine learning algorithms that are sensitive to the target scale. 

The important hyperparameter introduced above is $\tau$. Essentially, it is a window size for which the information about the future player performance is aggregated. Small values of the hyperparameter like 1 $s$ lead to noisy target values, while large values like 10 minutes neglect the subtle yet important changes in the player performance.
$\tau$ is commonly referred to as a \textit{forecasting horizon}.

% Below we'll refer to this hyperparameter as \textit{window size}.

$p_{\tau}(t)$ is a well defined metric for evaluating the player effectiveness, and it is possible to predict this value directly, thus considering the problem as a regression problem.
However, it is unclear how to interpret the quality of regression results in an understandable and interpretable way.
Formulating the problem in classification terms helps measuring the quality of prediction by more comprehensive classification metrics like accuracy, ROC AUC, and others. These metrics are much easier to compare with results obtained on other data or by other models.

The natural way to claim if a person plays well or bad at the moment is to compare the current performance with his or her average performance in the past.
It is important to consider the past events only to avoid overfitting.
% Then, for each 
Formally:
\begin{gather}
y_{\tau}(t) = [p_{\tau}(t) > \overline{p_{\tau}(t)}],\\
\overline{p_{\tau}(t)} = \frac{\sum_{t' < t}{p_{\tau}(t')}}{\sum_{t' < t}1},
\end{gather}
where $y_{\tau}(t)\in \{0,1\}$ and equals 1 in case of good game performance and 0 otherwise; $\overline{p_{\tau}(t)}$ is an average player performance in the past.% $\tau$ seconds.

\begin{figure*}[h]
    \centering
    \includegraphics[width=0.8\linewidth]{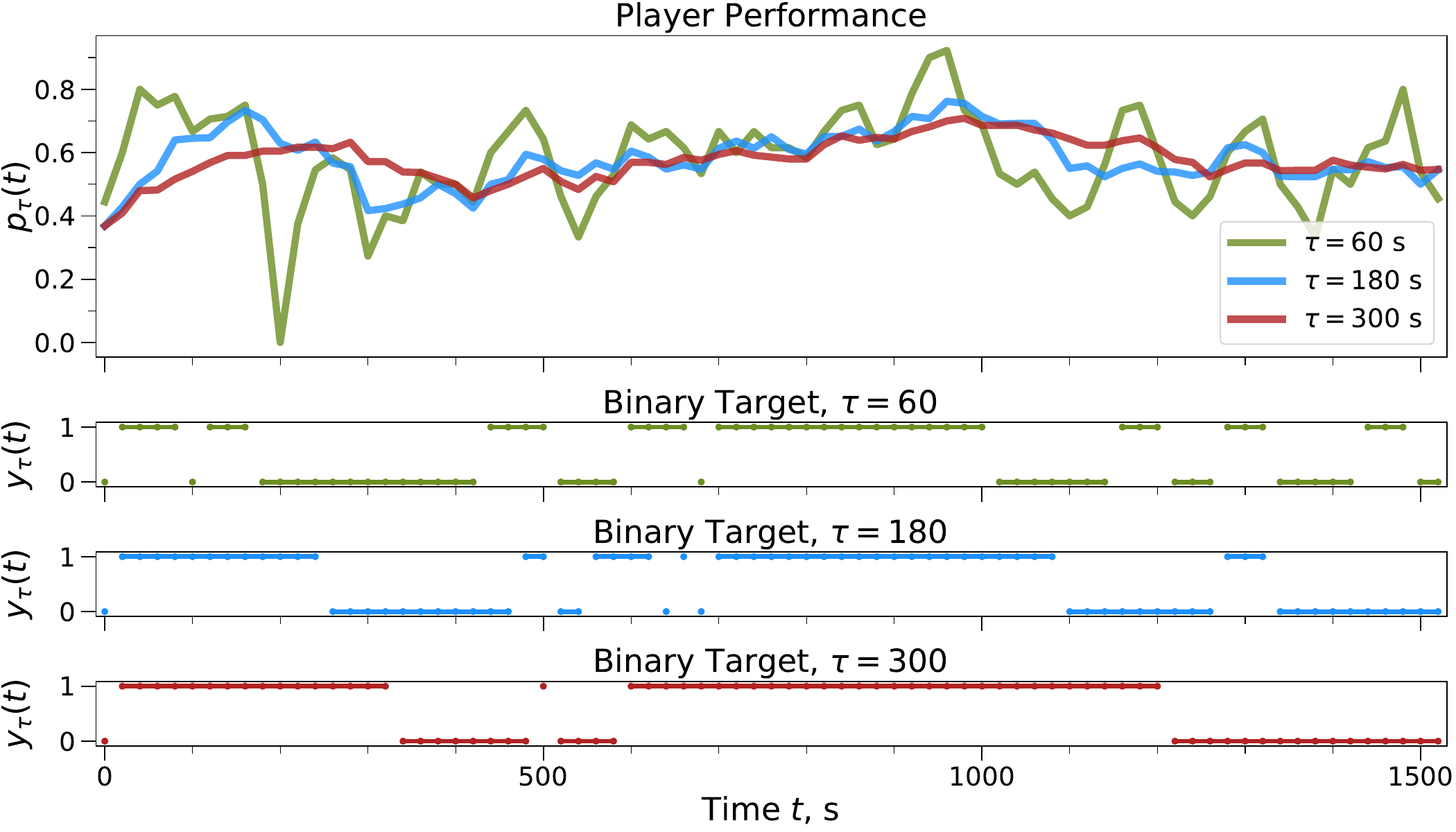}
    \caption{Player performance as the kills proportion $p_{\tau}(t)$ and its binarization for three forecasting horizons $\tau$ for a player in a dataset.
    % for a 
    % Kills ratio samples for three forecasting horizons for one player.
    }
    \label{target}
\end{figure*}

Figure~\ref{target} demonstrates how the kills ratio $p_{\tau}(t)$ and corresponding binary target $\yy(t)$ changes over time for three forecasting horizons.

The substantial advantage of using $y_{\tau}(t)$ instead of $p_{\tau}(t)$
is target unification between the players. $0$ and $1$ values of $y_{\tau}(t)$ have the same meaning for all players and imply bad and good current performance, respectively.
That is not the case for raw values of $p_{\tau}(t)$, because the same values of $p_{\tau}(t)$ may be good for one player, but bad for another one. For example, 0.5 kills ratio may be
an achievement
% a very good value
for a newbie player, but a failure for a professional player.

Another justification of using $y_{\tau}(t)$ as a target is robustness to the skill of other players on a server. Player's score $p_{\tau}(t)$ may be too low in absolute value because of the strong opponents, but target $y_{\tau}(t)$ is robust because it evaluates the performance within one game.
The motivation for this target is to unify the target variable for all players and to provide in advance an immediate feedback for a player or for a manager that something is going wrong.

Predicting future player performance $y_{\tau}(t)$ is essential for coaches and progressing players as it provides the quick feedback on players' actions.
That might help identify the inevitable failures in player performance, e.g. burnout, fatigue, etc., in advance and take measures to help the player to recover or even to change the player during the  eSports competition.
This target is also helpful for learning purposes:  although a person plays very well or very poor, it helps find the moments when the player performs a bit better or worse than average.

Despite we formulate the performance metric in terms of kills/deaths, the metric is directly applicable to the majority of First-Player Shooters, as well as other games including kills/deaths. The performance metrics for these and other games can be calculated in other terms (such as gold, scores, progress, etc.), while the data processing and algorithms may be the same.

% And also as a learning tool

% $p_{\tau}(t)$ is a nice metric for evaluating effectiveness, but it contains no information about performance dynamics. The latter can be estimated as

% \begin{equation}
%     \Delta p_{\tau}(t) = p_{\tau}(t+\tau) p_{\tau}(t) \label{delta}
% \end{equation}
% and basically equals the difference of player perfomance before and after moment $t$, or, in other words, performance dynamics.

% We decided to predict a binary value whether player performance has a positive dynamics or not. That's a classification problem, so we defined a positive class 1 for positive dynamics and negative class 0 for negative dynamics. 

% The motivation for this target is to unify target variable for all players and to provide an immediate feedback for a player or a coach that something is going wrong in advance.

% \section{Data Analysis}
\subsection{Predictive models} \label{alg_section}

We trained four models for predicting a player performance using the data from sensors: baseline model, logistic regression, recurrent neural network, and recurrent neural network with attention. In this section, we describe these methods in detail.
The output for all models is the probability that the person will play better in some fixed period in the future.
All the models are evaluated by ROC AUC score discussed in section~\ref{metric}.

% \begin{enumerate}
%     \item Baseline. model predicting current player .
%     \item Logistic Regression.
%     \item Neural Network.
%     \item Neural Network with Attention. 
% \end{enumerate}

\subsubsection{Baseline}

Before training a complex model, it is crucial to set up a simple baseline to compare it with.
A common practice in time series analysis
is to establish a baseline model using a current target value as a future prediction. For our problem, the baseline uses average player performance in the last $\tau$ seconds as a prediction. In other words, baseline prediction is $p_{\tau}(t)$. This prediction is correct because $p_{\tau}(t)$ takes values from $0$ to $1$, and then they treated as probabilities
by the algorithm used to calculate the ROC AUC metric.
% by ROC AUC scorer.
% when ROC AUC score being calculated.

\subsubsection{Logistic Regression}

Logistic regression~\cite{logistic_regression} is a simple and robust linear classification algorithm. It takes a feature vector $\xx(t) \in \RRR^n$ as an  input at the moment of time $t$ and provides the probability $y(t)$ equals $1$ as an output:
\begin{gather}
P(\yy(t) = 1|\xx(t)) = \frac1{1 + \exp(-\langle \ww, \xx(t) \rangle + b)},
\end{gather}
where $\ww \in \RRR^n$ is the learnable weights vector, $b \in \RRR$ is the bias term. In our study, the dimensionality is $n=15$ since we used 15 values from sensors.
Logistic regression can capture only linear dependencies in the data because the feature vector $\xx(t)$ involved in dot product with vector $\ww$ only.

% Logistic regression prediction is connected with distance from current sensors measurements $x_t$ to a
% hyperplane given by $\langle w, x \rangle = b$.

% In our problem 

% \subsection{Classical Machine Learning Models}

%  We used three classical machine learning alrogithms for this purpose.
% % In order to check the possibility to predict player performance dynamics during the game we built several machine learning models.

% \begin{enumerate}
%     \item \textit{Logistic regression}. The simple and robust linear classification algorithm \cite{logistic_regression},
%     \item \textit{Random forest}. An ensemble of diverse decision trees voting for the optimal class \cite{random_forest}. We used 100 estimators with maximum tree depth 3,
%     \item \textit{Support vector machine}. Nonlinear method for data separation in a hidden space \cite{svm}. We used radial basis functions (RBF) kernel.
% \end{enumerate}

% \subsection{Deep learning approach}
\subsubsection{Recurrent Neural Network}

A neural network can be considered as a nonlinear generalization of logistic regression.
In this subsection, we first describe the essential components used in the network and then describe the entire architecture.

% \subsubsubsection{Recurrent Neural Network Background} \label{rnn_section}
\textbf{Recurrent Neural Network Background} \label{rnn_section}

A Recurrent Neural Network (RNN) is a network that
% maintains some state.
maintains an internal state inherent to
some sequence of events.
It is proven to be efficient for discrete time series prediction \cite{rnn_review}.
One of the simplest examples of RNN is a neural network with one hidden layer.

Denote the sequence of input features and targets as
% $\big\{\xx(0), \xx(1), \dots, \xx(T-1)\big\}$
$\Big\{\xx\big(0\big), \xx\big(\dt\big), \dots, \xx\big((N-1)\dt\big)\Big\}$
and
% $\big\{y(0), y(1), \dots, y(T-1)\big\}$
$\Big\{y\big(0\big), y\big(\dt\big), \dots, y\big((N-1)\dt\big)\Big\}$
respectively, where $\xx(t) \in \RRR^n$ is the  $n$-dimensional feature vector for the moment $t$, $y(t)\in \RRR$ is the corresponding target, $\dt$ is the time step used for discretization, $N\in\NNN$ is the total number of steps.

At each moment $t$, the recurrent network has an $m$-dimensional hidden state vector $\hh(t) \in \RRR^m$ which is calculated
using the current input $\xx(t)$ and the previous hidden state $\hh(t-\dt)$:
% using the input $\xx(t)$ at the current time step and the hidden state from the previous step $\hh(t-\dt)$:
\begin{equation}
    \hh(t) = \tanh \Big(\mathbf{W} \xx(t) + \mathbf{U} \hh(t-\dt) + \mathbf{b} \Big),
\end{equation}
% where $\hh(t) \in \RRR^{m}$ is a hidden state for the step $t$;
where $\mathbf{W} \in \RRR^{m \times n}$, $\mathbf{U} \in \RRR^{m \times m}$, $\mathbf{b} \in \RRR^{m}$ are the learnable matrices and the bias vector.

The intuition behind using the RNN architecture might be in considering the hidden state of the network as a current state of the player represented by the data from sensors. The state is a vector with many components and some of them can present how well the person will play. The final prediction $\hat{y}_\tau(t)$ at the moment $t$ is calculated by the feed-forward network consisting of $1$ or more linear layers:
\begin{gather}
\hat{y}_\tau(t) = f(\hh(t)),
\end{gather}
where $f:\RRR^m \to \RRR$ is the function corresponding to the feed-forward network. We used a sigmoid function as a final activation for the network to ensure $\hat{y}_\tau(t) \in [0,1]$, so $\hat{y}_\tau(t)$ has a meaning of probability.

% \subsubsubsection{Gated Recurrent Unit} \label{gru_section}
\textbf{Gated Recurrent Unit} \label{gru_section}

More advanced modifications of the recurrent layer include the Gated Recurrent Unit (GRU)~\cite{gru_paper} and Long Short-Term Memory (LSTM)~\cite{LSTM}. Both of them utilize the gating mechanism to better control the flow of  information. LSTM architecture incorporates an   input, output, and forget gates and a memory cell, while a simpler GRU architecture uses the update and reset gates only.
We found GRU performs better in our task,
so we formally define it as follows:
% It is defined as:
\begin{gather}
\hh(t) = (1 - \zz(t)) \odot \hh(t-\dt) + \zz(t) \odot \tilde{\hh}(t),\\
\tilde{\hh}(t) = \tanh(\WW_h \xx(t) + \UU_h (\rr(t) \odot \hh(t-\dt)) + \bb_h),\\
\zz(t) = \sigma(\WW_z \xx(t) + \UU_z \hh(t-\dt) + \bb_z),\\
\rr(t) = \sigma(\WW_r \xx(t) + \UU_r \hh(t-\dt) + \bb_r),
\end{gather}
where $\zz(t)$ and $\rr(t)$ are the update and reset gates, $\WW_h, \WW_z, \WW_r, \UU_h, \UU_z, \UU_r, \bb_h, \bb_z, \bb_r$ are the learnable parameters, $\odot$ is Hadamard product, $\sigma$ is the sigmoid function.
% The intuition behind TODO: please describe

% LSTM architecture defined similarly, but it also has an input gate and a memory cell.

% \subsection{Recurrent neural network with attention}

% \subsubsubsection{Attention Mechanism} \label{att_section}
\textbf{Attention Mechanism} \label{att_section}

A popular technique for improving the network quality and interpretability is an attention mechanism. 
%Attention can be applied to temporal data about previous hidden, in
Temporal attention can help emphasize the relevant hidden states from the past. Input attention helps to select the essential input features. It is also possible to combine both of them~ \cite{input_attention}. Since the proposed GRU model uses only one previous hidden state for prediction, it makes no sense to use the temporal attention. However, the input attention can be used.

The attention layer provides the weights vector $\aa(t)\in\RRR^n$, which is applied to a vector $\xx'(t)\in\RRR^n$
by the element-wise multiplication:

\begin{gather}
    \tilde{\xx}(t) = \xx'(t) \odot \aa(t).
\end{gather}

This operation demonstrates important components in the vector $\xx(t)$ while decreasing the contribution of its non-relevant components. Typically, $\aa(t)$ components are bounded by $0$ and $1$ and
% Commonly, $\aa(t)$ is
produced by another linear layer integrated into the network. In order to consider both the current and the previous data from sensors, the attention layer takes current measurements $\xx(t)$ and hidden state $\hh(t)$ produced by GRU:
\begin{gather}
\aa(t) = \sigma(\WW_a [\hh(t-\dt), \xx(t)] + \bb_a),
\end{gather}
where the $\WW_a \in \RRR^{(m + n) \times m}, \bb_a \in \RRR^{m}$ are trainable parameters.

The intuition behind the input attention mechanism is a feature selection. Based on the current input and hidden state, it can help ignoring the  uninformative features for each moment of time and keep relevant features unaltered. The ability to provide the time-dependent feature importance is a significant advantage compared to other methods, which can either provide feature importance for one particular moment or provide it only for the whole time series.% in general.

\textbf{Recurrent Neural Network With Attention Architecture}  \label{rnn_att_arch}

\begin{figure*}[t]
    \centering
    \includegraphics[width=0.85\linewidth]{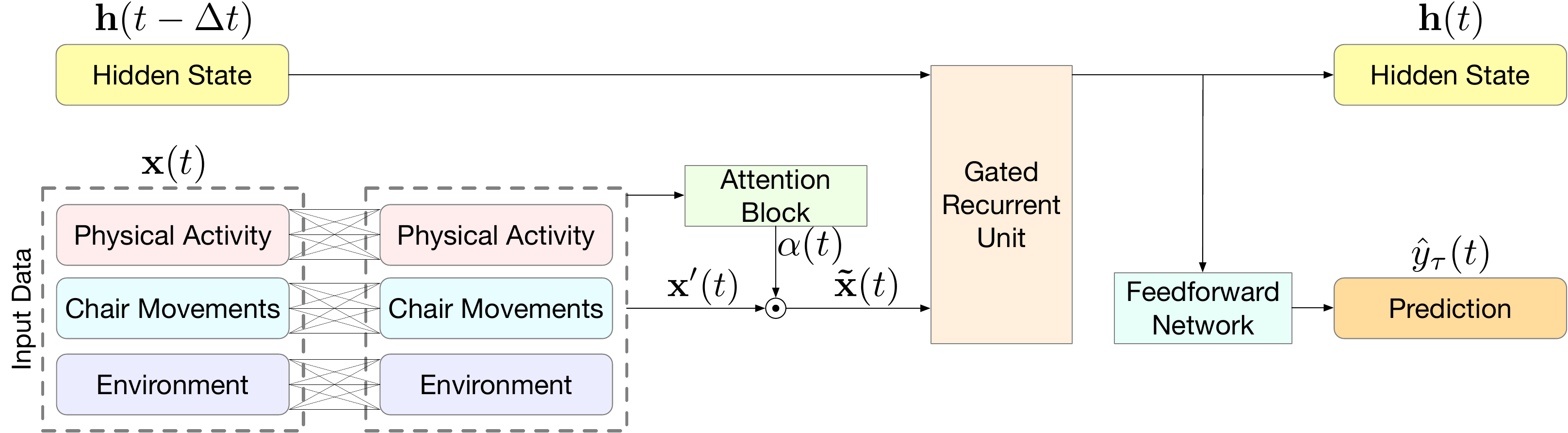}
    \caption{Recurrent neural network architecture.}
    \label{rnn_arch}
\end{figure*}

% Based on details provided in \ref{rnn_section}, \ref{gru_section} and \ref{att_section} the neural network architecture can be easily visualized in Figure~\ref{rnn_arch}. 

The neural network architecture is shown in Figure~\ref{rnn_arch}. First, the sensor data are processed by the dense layers for each feature group. Then the attention block is applied
% to pay attention to
to amplify a signal from the
features relevant at the moment. The resulting vector goes through the GRU cell to update the hidden state $\hh(t)$. This hidden state is saved for further iterations and goes through
a feed-forward network to form the eventual prediction $\hat{\yy}(t)$. The total inference time is about 5 $ms$ on a CPU.

The network was trained by the 
truncated backpropagation through time \cite{tbptt} technique designed for RNNs and Adam optimizer~ \cite{adam_paper}
with learning rate warmup \cite{warmup} technique to improve the convergence.
For attention and feed-forward networks, we used $1$ and $2$ linear layers, respectively, with ReLU nonlinearity. This activation function can improve the convergence and numerical stability~\cite{relu}. To validate whether the attention mechanism helps  improve network performance, we also trained another network without the attention block.

% First of all, input and hidden state go through attention network, then attention weights $\aa$ applied to the input vector and result goes to RNN. After RNN step we have an updated hidden vector $\hh(t+1)$, which is saved for further iterations and also goes through feed-forward network to form a prediction $\mathbf{y}(t)$.

GRU cell is a crucial part of the network architecture. 
% Because of 
It helps accumulate information about previous player states, so the network can use the retrospective context for prediction. This information is stored in the hidden layer of the GRU cell.
% We found that 8 neurons in the hidden layer works for our problem the best.
According to the experiments, 8 neurons in the hidden layer works for our problem the best. Too few neurons caused low predictive power, while too many led to overfitting.

The motivation to use separate linear layers for three feature groups is to combine more complex features from the sensors data and to preserve the  disentangled feature representation for the attention layer. This is an analog of grouped convolutions~\cite{grouped_layers} in convolutional networks. For the same reasons, we applied the  attention to each feature group separately, thus having a 3-dimensional attention vector at each moment of time.

% It consist of attention network block for estimating attention weights $\aa$, recurrent neural network and f

\subsection{Validation} 
\label{evaluation_section}

\subsubsection{Training and Evaluation Process}
In order to correctly estimate the generalization capabilities of classical machine learning algorithms, we used repeated cross-validation \cite{repeated_cross_validation}. In particular, we randomly split 21 players into the train and test groups of 16 and 5 players, respectively. Then we  trained the algorithms and repeated this procedure  again 100 times. It helps to lower the variance in evaluation.

For neural networks, we also used a validation set, so we randomly split players into train, validation, and test sets with 11, 5, and 5 players, respectively.
% players were randomly splitted to sets of 11, 5, 5 people for train, validation and test respectively.
The neural network has been trained until the error on the validation set is not improving for 5 epochs (early stopping procedure).
One training epoch comprises of 20 batches. Each batch consists of input features and targets for all the time steps for a randomly selected player from the train set.
To minimize the randomness in the evaluation results, each network has been trained 15 times with random weight initializations and train/val/test split.
In both cases, the input features for train, test, and validation sets are normalized based on the 
mean values and standard deviations calculated on the train set.

\subsubsection{Evaluation Metric} 
\label{metric}

Due to construction, the target is balanced: 50.1\% belong to the positive class and 49.9\% to the negative class.
The common metric for classification evaluation is the area under the receiver operating characteristic curve (ROC AUC) \cite{roc_auc}. It ranges from $0$ to $1$ with the $0.5$ score for random guessing. Higher values are better.

For proper evaluation, we first calculated the ROC AUC scores for each individual participant in the train, validation, or test sets, and then averaged the results. That is the proper evaluation because it estimates how well the model can separate the high/low performance conditions for one individual participant and does not benefit from separating the  participants between themselves in the case when the metric is calculated on predictions for all the  participants.

% , not for the entire set
% doesn't reward model to distingu
% forces model to properly 
% That evaluation 
% We averaged AUC scores

\section{Results} 
\label{results_section}

\subsection{Performance of Algorithms}
We have trained the predictive models for several time steps $\dt$ (see Section \ref{time_series_resampling}) and forecasting horizons $\tau$ (see Section \ref{player_performance_evaluation}) combinations. According to our experiments, reasonable ranges are from 5 to 30 $s$ for time step, and from 60 to 300 $s$ for the forecasting horizon. The average results for the neural network with attention are shown in Table~\ref{nn_att_results}.

\begin{table}[t]
\caption{Experimental results for the neural network with the attention for varying time steps and forecasting horizons. Results are reported in ROC AUC score.}
\label{nn_att_results}
% \begin{tabular}{|p{1.55cm}|c|c|c|c|c|c|}
\begin{tabular}{|c|c|c|c|c|c|c|}
\hline
\multirow{2}{1.55cm}{\centering Forecasting Horizon $\tau$, s} & \multicolumn{6}{|c|}{Time Step $\dt$, s} \\ \cline{2-7}
 & 5 & 10 & 15 & 20 & 24 & 30 \\ \hline
% 60 & 60.2 & 65.8 & 63.2 & 61.7 & 64.5 & \textbf{68.9} \\ \hline
% 120 & 60.9 & 62.1 & \textbf{68.2} & 63.1 & 68.0 & 65.7 \\ \hline
% 180 & 61.1 & 62.2 & 65.2 & \textbf{71.1} & 68.2 & 64.2 \\ \hline
% 240 & 66.7 & 66.1 & 65.4 & \textbf{72.6} & 72.4 & 70.3 \\ \hline
% 300 & 61.9 & 66.7 & 72.1 & 68.0 & 71.7 & \textbf{72.4} \\ \hline
60 & 0.602 & 0.658 & 0.632 & 0.617 & 0.645 & \textbf{0.689} \\ \hline
120 & 0.609 & 0.621 & \textbf{0.682} & 0.631 & 0.680 & 0.657 \\ \hline
180 & 0.611 & 0.622 & 0.652 & \textbf{0.711} & 0.682 & 0.642 \\ \hline
240 & 0.667 & 0.661 & 0.654 & \textbf{0.726} & 0.724 & 0.703 \\ \hline
300 & 0.619 & 0.667 & 0.721 & 0.680 & 0.717 & \textbf{0.724} \\ \hline
\end{tabular}
\end{table}

According to Table~\ref{nn_att_results}, the best time step value is about 20 $s$, and the best forecasting horizon for the model is from 3 to 5 minutes.
In other words,
% model has the best predictive power
the optimal way to predict the player behavior is to aggregate the sensors data every 20 $s$ and make a prediction for the next 3-5 minutes.

\begin{figure}[t]
    \includegraphics[width=1.0\linewidth]{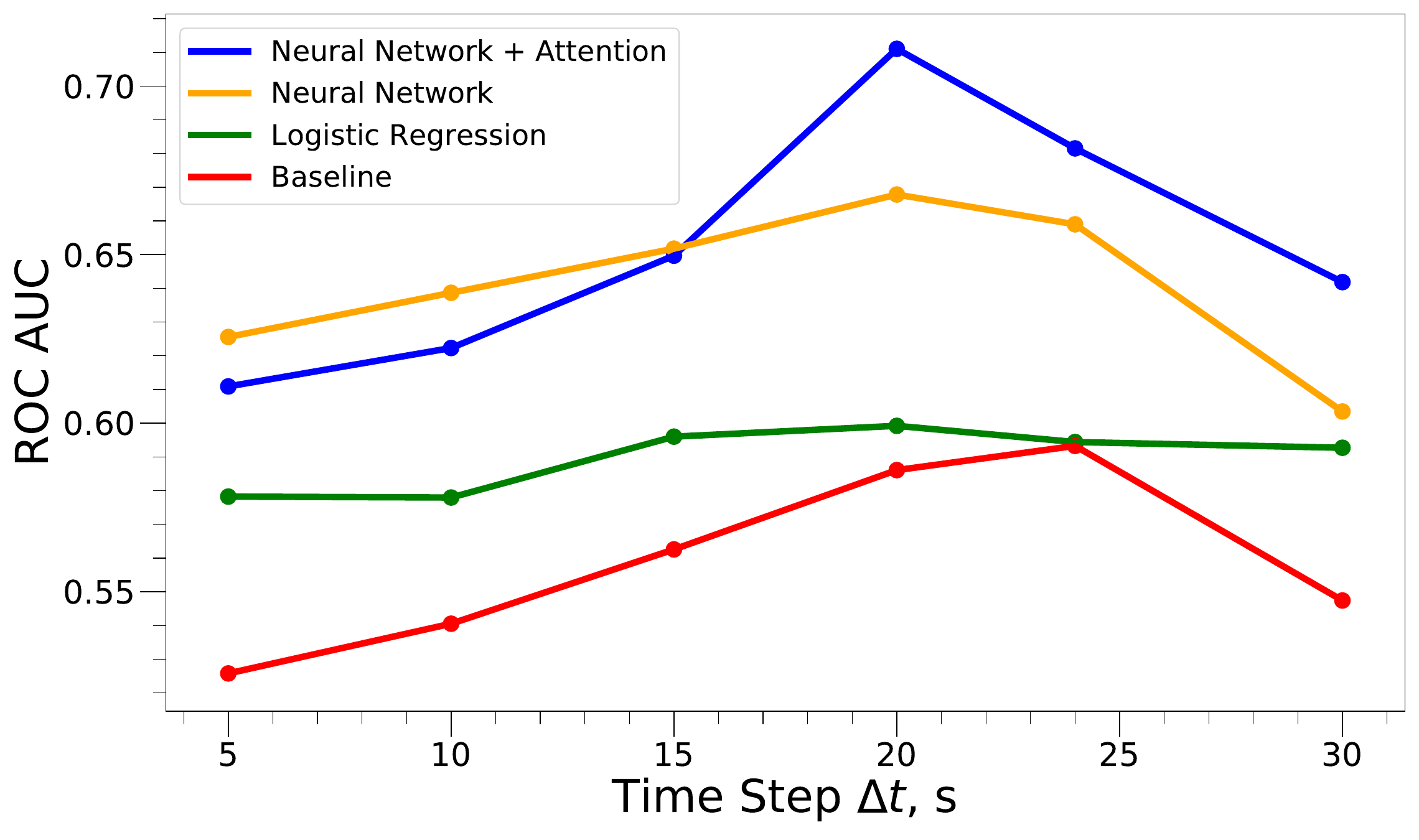}
    \caption{Algorithms performance w.r.t. discretization time step $\dt$ with fixed forecasting horizon $\tau=180$ $s$.}
    \label{time_step_figure}
\end{figure}

We have compared the performance of the algorithms described in Section~\ref{alg_section} with respect to the time step values with fixed forecasting horizon $\tau$ equal 180 $s$. The results are shown in Figure~\ref{time_step_figure}. There is a clear peak near 20-25 $s$ for all the methods. The neural network consistently outperforms the logistic regression and the baseline model. The use of the attention block helps increase the model score.

% Algorithms perfomance w.r.t. to time step discretization value are shown in Figure \ref{time_step_figure}. The window size for player performance evaluation here is fixed and equals 5 minutes.  In general, algorithms perform better with increasing time step. However, comparing results with time step 40 and more is not correct, in this case some players are missing in the data because their gaming sessions has too few time steps.

\begin{figure}[t]
    \includegraphics[width=1.0\linewidth]{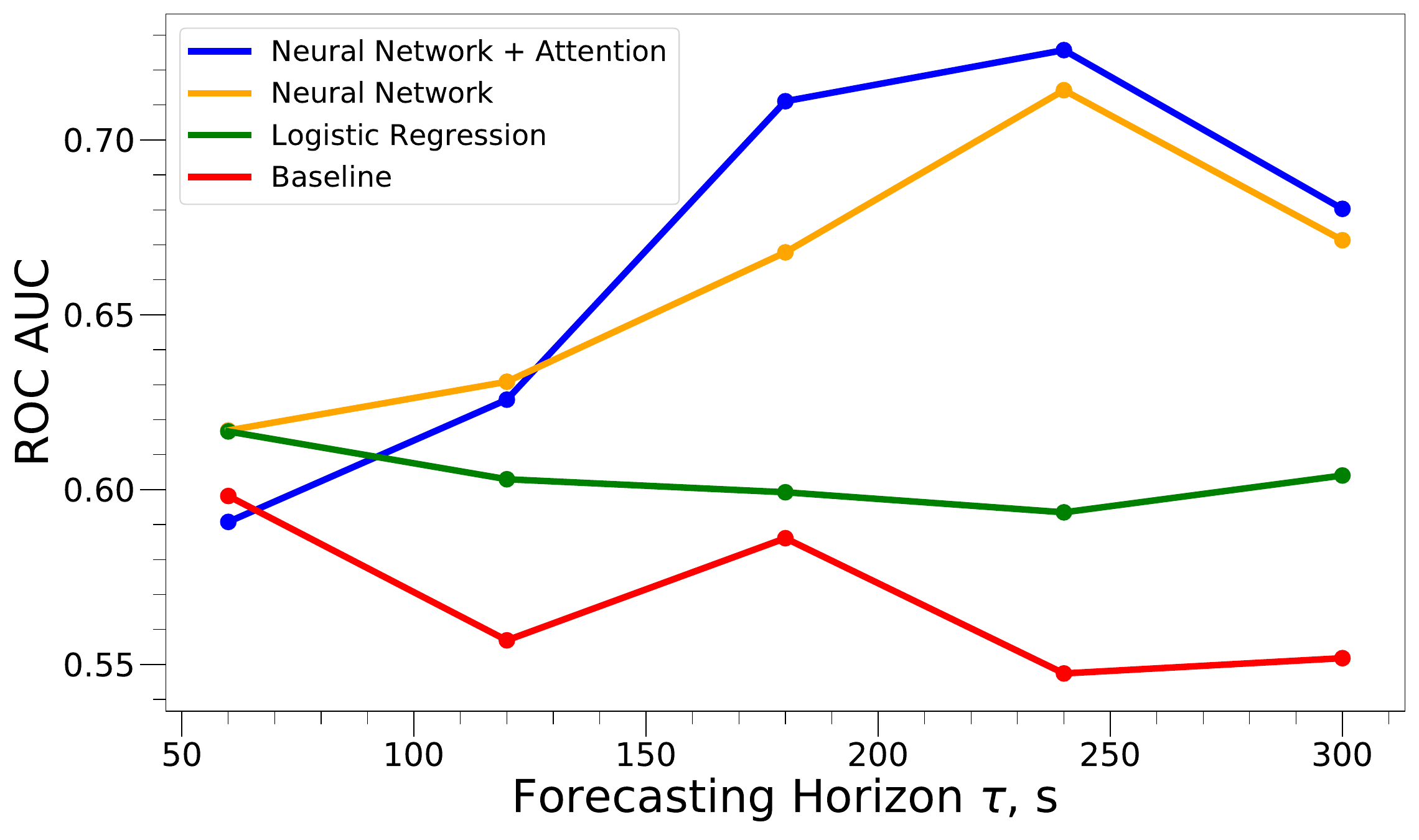}
    % \caption{Algorithms performance w.r.t. \textit{forecasting horizon}.}
    \caption{Algorithms performance w.r.t. forecasting horizon $\tau$ with fixed time step $\dt=20$ $s$.}
    \label{window_size_figure}
\end{figure}

% In order to estimate algorithms perfomance first of all we need to set values for hyperparameter \textit{window size} (see section \ref{player_performance_evaluation}), which corresponds to window size for averaging player performance, and for hyperparameter \textit{time step} (see section \ref{time_series_resampling}, which sets up the sampling rate.

Figure~\ref{window_size_figure} demonstrates the relation between the forecasting horizon and algorithms performance with the time step $\dt$ equal to 20 $s$. The neural network outperforms other methods and achieves the maximum performance for forecasting horizons in the range from 3 to 5 minutes. The attention block helps improve the model.

% From figures \ref{window_size_figure} and \ref{time_step_figure} it clearly follows that neural networks outperform classical machine learning algorithms in the problem. We'll focus on them further.

\subsection{Feature importance}
\label{feature_importance_section}

\begin{figure*}[h]%{0.5\textwidth}
    \includegraphics[width=\linewidth]{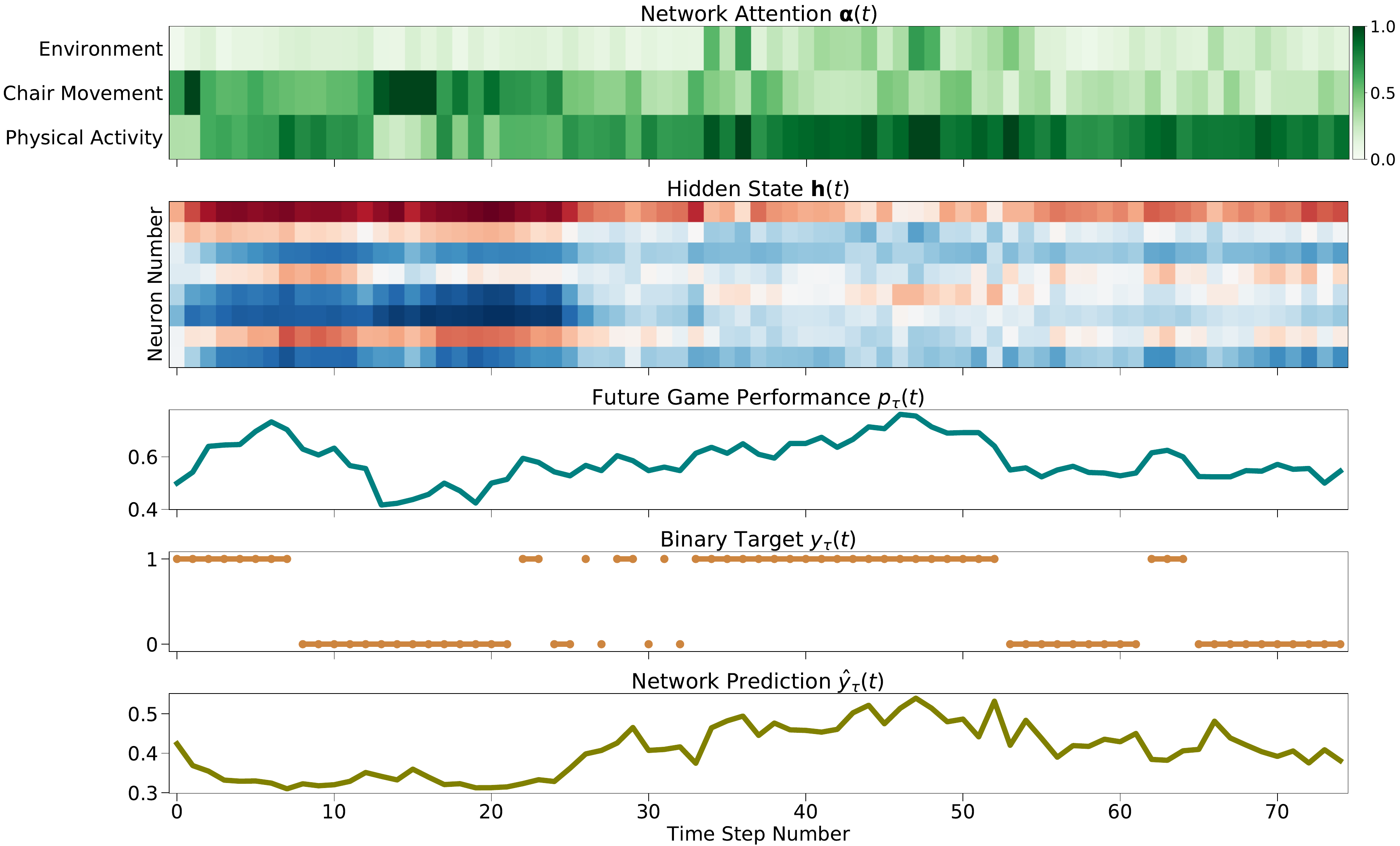}
    % \caption{Sample attention visualization for a player in a test set.}
    \caption{Visualization of attention weights $\aa(t)$, hidden state $\hh(t)$, game performance $p_\tau(t)$, binary target $\yy(t)$ and network prediction $\hat{y}_\tau(t)$
    w.r.t. time step number
    for a player in the test set.}
    \label{att_visualization_0}
\end{figure*}

In order to interpret the neural network predictions, we calculated the feature importances and visualized predictions of a pretrained network and its internal state for the discretization step equal to 20 $s$ and the forecasting horizon equal to  3 minutes.
Figure~\ref{att_visualization_0} shows how attention, network hidden state, target, and network prediction change over time.
Clearly, the importance of different features varies over time, and periodically the data from some sensors is non-relevant.
The network hidden state, which can be treated as a player state, also varies during the game. 

% In Section~\ref{results_section} we investigated that network
% In this section we visualize its predictions, internal state and

% It's interesting to visualize which sensors contribure to prediction the most in different periods of time. We selected RNN with sigmoid attention function as an illustrative example, and fitted it on a random train set using early stopping on a validation set. Then we run the network for all players and logged input attentions for all time steps. 

% In the Figures~\ref{att_visualization_1} and \ref{att_visualization_2} we demonstrate how input attention and hidden state of RNN change during the game. We also added player performance dynamics, corresponding target and model predict below.

% It's clear that some sensors are non-relevant during specific periods of time. For example, for both players data about $x$-component of acceleration from a chair is not relevant most of the time. Mouse movements in Figure \label{att_visualization_2} are mostly used in the beginning of the match, but become less valuable later.

% Interestingly, the attention weights as well as hidden state of the player can change dramatically during the game session. For both players there are breakpoints that also correspond to a change in performance dynamics shown below.

\begin{figure}[]
		\centering
		\includegraphics[width=0.9\linewidth]{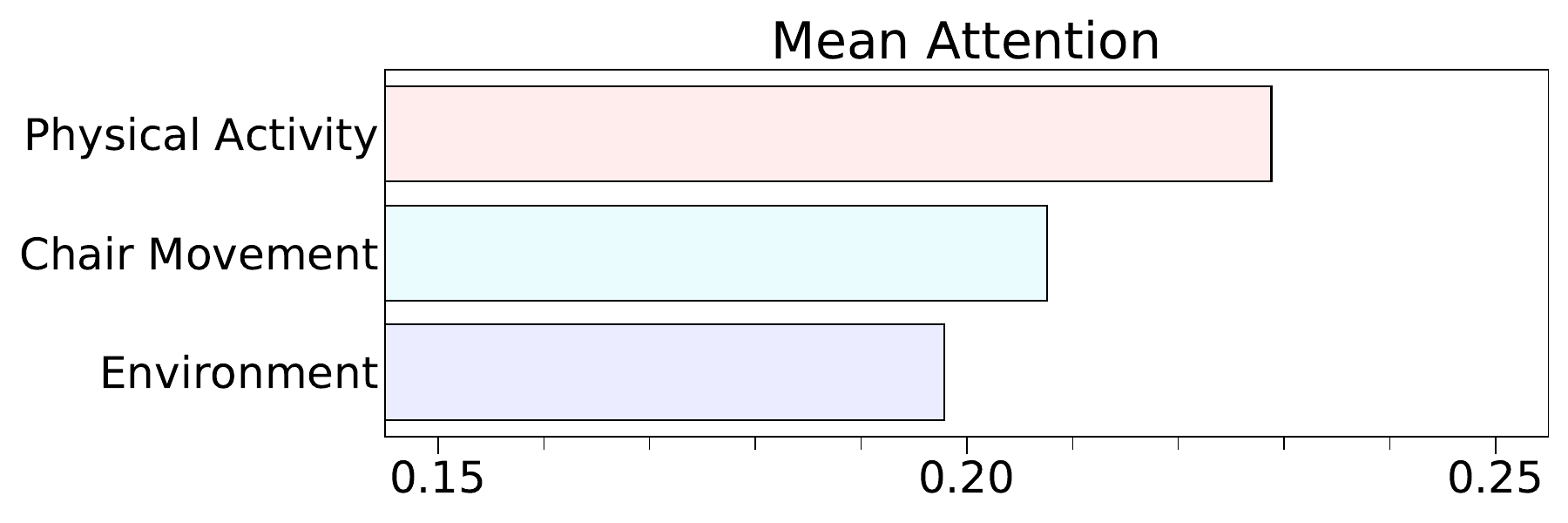}
        \captionof{figure}{Feature importance as the  mean neural network attention.}
        \label{feature_importance_fig}
\end{figure}

To calculate the feature importance, we trained 100 instances of the neural network with random weight initializations and the train/val/test splits, and the averaged attentions on the test set for the best epoch of each neural network. Afterwards, we averaged the results across all networks. The results are shown in
Table~\ref{feature_importance_table}
and
Figure~\ref{feature_importance_fig}.

\begin{table}[b]
	%{0.5\linewidth}
		\caption{Mean neural network attention for feature groups.}
% 		\caption{Feature importance as a mean neural network attention.}
        \label{feature_importance_table}
        \centering
        \begin{tabular}{|l|c|}
        \hline
        \textbf{Feature Group} & \textbf{Mean Attention} \\ \hline
        Physiological Activity & 0.228 \\ \hline
        Chair Movement & 0.207 \\ \hline
        Environment & 0.197 \\ \hline
        \end{tabular}
	\end{table}\hfill

Information about physiological activity such as heart rate, muscle activity, hand movement, and so on is the most relevant for the network.
It is worth noting that all the feature groups 
have considerable importance,
thus
% and
contribute to the overall prediction. Since the  features in each feature group are mixed,
we can not estimate the feature importance for every raw feature.

\subsection{Discussion} \label{discussion_section}

Experimental results have shown practical feasibility of predicting the eSports players behavior using only data from sensors. 
The system updates the prediction several times per minute, so
% and
this interactivity is enough for potential users like eSports managers or
% training
professional
players to understand that something goes wrong or to get a quick feedback. % almost in real time.
Feedback from the algorithm about the performance prediction and feature importance may suggest the  users to change their gaming behavior. For example, an eSports manager may understand in time that bad  results of the team are connected with the stuffy environmental conditions and adjust the air conditioning. % Any other example?
Users may find information from the system useful to
make the radical decisions, e.g., changing a player or gaming equipment (computer mouse, display, game chair, etc.).

% The following potential actions
% % fast changes
% % fundamental changes
% % radical changes

We have found that 20 $s$ time step and 3-5 minutes behavior forecasting horizon
% the best
are the most natural
parameters for CS:GO, but potential users can set up any other hyperparameters depending on their scenario. 
Significant advantage caused by the diversified training dataset is universality for the player skill level enabling a wide range of potential users.
Negligible model inference time and data collection time on a PC
proves the principal feasibility of deploying the model on the edge devices, some of which might be designed specifically for neural network operations.
However, the model retraining still requires high computational capability. 
Larger and more diversified dataset could improve the performance of prediction. 
%At the same time it is our future work to scale the proposed solution to other video games. 

% Concrete usage
% Indeed, sensing system can be adapted for other games or domains
% Maybe mention Other games here

% The data collection delay and inference time for algorithms are

% That also proves the possibility of tratata edge devices.
% The algorithms also doesn't require much memory tratata edge devices.
% The network doesn't have to be fitted for each new player
% scalability

\subsection{Limitations}

The limitations of the study is a small number of participants involved, the fuzziness of the definition of player's performance, and overrepresentation of data from males and young people in the dataset.
Future work includes more diverse and extensive data
collection with more subjects recorded, and the investigation of better metrics of players' performance. These would allow researchers to utilize
more complex machine learning methods and to develop more reliable and robust models.

\section{Conclusions} 
\label{conclusions_section}

In this article, we have reported on the AI-enabled system for predicting the performance of eSports players using only the data from heterogeneous sensors. 
The system consists of a number of sensors capable of recording players' physiological data, movements on the game chair, and environmental conditions.
Upon data collection we have processed them into time series with meaningful sensors features and the  target extracted from the game events. 
%including a formula for estimating the player performance. 
The Recurrent Neural Network (RNN) demonstrated the best performance comparing to baseline and logistic regression. Application of the attention mechanism for RNN has helped to interpret the network predictions as well as to extract the feature importances. 
Our work showed the connection between the player performance and the data from sensors
as well as the possibility of making a real-time system for training and forecasting in eSports.

We have also investigated potential applications of the proposed AI system in the eSports domain. Given the growth of eSports activity due to the coronavirus pandemic in 2019--2021 and rapid development of consumer wearable devices within the last years, this work 
% establish
shows the prospectives of
full-fledged research in the intersection of these two fields. Moreover, the model trained on the eSports domain can be transferred to other domains using domain adaptation methods
% \cite{domain_adaptation}
to estimate user's performance similar to estimating in-game performance.

%Future work includes extensive data collection with more participants involved, longer game sessions, and other games played. Lack of data collected is currently the major bottleneck for understanding players' behavior. More data allow researchers to train more generalizable and complex models as well as to test new hypotheses.

Considering hunders of millions of active gamers in the world and widespread of wearables, crowdsourcing data collection 
%\cite{crowdsourcing_data_collection} 
is a promising way to collect the data on a global scale. We also see potential improvements in our system with computer vision methods. Emotion recognition and pose estimation techniques applied to data collected from web camera can provide more information about the current state of a player.

% % Raw data were processed by the resampling techniques and further analyzed by the machine learning algorithms.
% An important part of the work is the estimation of eSports athlete performance in each moment of time and extracting its dynamics.
% It has turned out that the best algorithm for predicting player performance dynamics is recurrent neural network. We have also proposed the input attention meachanism to increase the  model generalization and interpretability. That allowed us to visualize the relevance of the features during the game session and calculate the feature importance for data from the sensors.

%In future work, we intend to collect larger and more diversified dataset, try new neural network architectures, and investigate player performance estimation in other games. At the same time, we plan to squeeze our machine learning algorithms into mobile devices and embedded systems for ensuring mobile eSports analytics, which is another quickly developing research field.


% \begin{acks}
\section*{Acknowledgment}
The reported study was funded by RFBR according to the research project No. 18-29-22077.

Authors would like to thank professional eSports team DreamEaters for fruitful discussions and data collection.
% \end{acks}


\bibliography{references}{}
\bibliographystyle{IEEEtran}
% \bibliographystyle{ACM-Reference-Format}
% \bibliography{references}

\end{document}